\documentclass[prd,nofootinbib,showpacs,preprint]{revtex4-1}
\usepackage[T1]{fontenc}
\usepackage{amsmath,amssymb}
\usepackage{epsfig}
\usepackage{url}
\usepackage{graphicx}
\usepackage[usenames,dvipsnames]{color}
\usepackage{slashed}
\usepackage[colorlinks,citecolor=blue]{hyperref}
\usepackage{color}
\usepackage{cancel}
\newcommand{\be}{\begin{equation}}
\newcommand{\ee}{\end{equation}}
\newcommand{\bea}{\begin{eqnarray}}
\newcommand{\eea}{\end{eqnarray}}

\newcommand{\newc}{\newcommand}
\newc{\bi}{\begin{itemize}}
\newc{\ei}{\end{itemize}}

\newc{\ra}{\rightarrow}
\newc{\sq}   {\mbox{$\wt{q}$}}
\newc{\msq}  {\mbox{$m_{\sq}$}}
\newc{\gl}   {\mbox{$\wt{g}$}}
\newc{\mgl}  {\mbox{$m_{\gl}$}}
\def \met  {\mbox{${E\!\!\!\!/_T}$}}
\def \me  {\mbox{${E\!\!\!\!/}$}}
\newc{\wt}{\widetilde}
\def \ifb{{\rm fb}^{-1}}
\def \mch{m_{H_{\nu}^{\pm}}}
\def \ch{H_{\nu}^{\pm}}
\def \mhh{m_{H_{\nu}}}
\def \hh{H_{\nu}}
\def \mah{m_{A_{\nu}}}
\def \ah{A_{\nu}}
\def \mhn{m_{N}}
\def \hn{N}
\begin{document}
\preprint{HIP-2017-35/TH, HRI-RECAPP-2017-015}
\title{Exploring collider aspects of a neutrinophilic Higgs doublet model in multilepton channels}
\author{Katri Huitu}
\email{ katri.huitu@helsinki.f}
\affiliation{Department of Physics, and Helsinki Institute of Physics, P. O. Box 64, FI-00014 University of Helsinki, Finland}
\author{Timo J. K\"{a}rkk\"{a}inen}
\email{timo.j.karkkainen@helsinki.f}
\affiliation{Department of Physics, and Helsinki Institute of Physics, P. O. Box 64, FI-00014 University of Helsinki, Finland}
\author{Subhadeep Mondal}
\email{subhadeep.mondal@helsinki.fi}
\affiliation{Department of Physics, and Helsinki Institute of Physics, P. O. Box 64, FI-00014 University of Helsinki, Finland}
\author{Santosh Kumar Rai}
\email{skrai@hri.res.in}
\affiliation{Regional Centre for Accelerator-based Particle Physics,\\
Harish-Chandra Research Institute, HBNI, Jhusi, Allahabad - 211019, India}
\begin{abstract}
We consider a neutrinophilic Higgs scenario where the Standard Model is extended by one additional Higgs doublet and three generations 
of singlet right-handed Majorana neutrinos. Light neutrino masses are generated through mixing with the heavy neutrinos via 
Type-I seesaw mechanism when the neutrinophilic Higgs gets a vacuum expectation value (VEV). The Dirac neutrino Yukawa coupling 
in this scenario can be sizable compared to those in the canonical Type-I seesaw mechanism owing to the small neutrinophilic Higgs 
VEV giving rise to interesting phenomenological consequences. We have explored various signal regions likely to provide a hint 
of such a scenario at the LHC as well as at future $e^+e^-$ colliders. 
We have also highlighted the 
consequences of light neutrino mass hierarchies in collider phenomenology that can complement 
the findings of neutrino oscillation experiments. 
\end{abstract}
\maketitle
\section{Introduction}
The discovery of the 125 GeV Higgs boson \cite{Aad:2012tfa,Chatrchyan:2012xdj} has been 
a remarkable achievement of the Large Hadron Collider (LHC). This has provided us  a
closure regarding the predictions of the Standard Model (SM). While 
our quest towards understanding the physics beyond the Standard Model (BSM) continues, 
the 13 TeV run of the LHC is expected to make a big impact both in terms of higher energy 
reach and better precision by accumulating huge amount of data at large luminosity. 
The enigma of non-zero neutrino mass has pushed the theorists as well as experimentalists 
to develop new theories and experimental techniques in order to establish the 
right theoretical pathway towards unveiling the true nature of neutrino mass generation. 
The neutrino oscillation experiments have established the fact that at least two of the 
three light neutrinos are massive, and that they have sizable mixing among themselves 
(for a review, see \cite{GonzalezGarcia:2007ib}). The SM, lacking any right-handed neutrinos, 
is unable to account for these phenomena. This has led to a plethora of scenarios leading 
to neutrino mass generation \cite{Minkowski:1977sc,Mohapatra:1979ia,GellMann:1980vs,
Yanagida:1979as,Glashow:1979nm,Schechter:1981cv,Schechter:1980gr,Weinberg:1979sa,Weinberg:1980bf}. 
As the resulting neutrino mass eigenstates may be either Dirac or Majorana type, both scenarios have 
potentially unique signatures \cite{Keung:1983uu,Datta:1993nm,Almeida:2000pz,Panella:2001wq,Han:2006ip,delAguila:2007qnc,
Huitu:2008gf,Atre:2009rg,Keung:1983uu,Datta:1993nm,Almeida:2000pz,Panella:2001wq,Han:2006ip,Chen:2011hc,Dev:2013wba,Das:2017hmg} 
in the collider experiments. The LHC collaborations have put significant effort to extract 
any possible information about such scenarios from the accumulated data and the null results so far 
have only been able to constrain the parameter space of various neutrino mass models \cite{Abulencia:2007rd,
Chatrchyan:2012fla,ATLAS:2012yoa,Khachatryan:2015gha,BhupalDev:2012zg,Deppisch:2015qwa}. 

In the post Higgs discovery LHC era, the true nature of the scalar sector remains another vital area of 
interest. The natural question that arises is whether the 125 GeV Higgs is the only scalar as predicted 
by the SM or other exotic scalars exist alongside, as predicted by various BSM theories including 
some of the neutrino mass models \cite{Schechter:1980gr,Magg:1980ut}. The measurements of 
couplings of the 125 GeV Higgs with known SM particles have so far been consistent with the SM 
predictions \cite{CMSATLAS:comb}. Thus, even if this Higgs boson were indeed part of a larger scalar 
sector, its mixing with the other states would be small. There are still enough uncertainties in these 
measurements to allow new exotic scalar multiplets. Unless the LHC observes some hint of a new 
scalar, our only hope lies in the precision measurements of the Higgs couplings in order to constrain the 
BSM physics scenarios. Meanwhile, there has been a long term interest in the simplest 
two-Higgs doublet models (2HDM) (for a review, see \cite{Branco:2011iw}) which are also strongly 
motivated by supersymmetric scenarios. A two-Higgs doublet model predicts the presence of two CP-even, 
one CP-odd and two charged Higgses, one of the CP-even Higgs states being the 125 GeV Higgs boson. 
Despite the presence of these additional scalar states, the mixing between the two doublets can be 
arranged so that the other scalars are practically decoupled from the SM Higgs. In such cases, the interaction 
of the SM-like Higgs with the exotic scalars may be so suppressed that any hint of such interactions 
can be very hard to pick up even with the precision measurements at the LHC. The hope of finding these scalars, 
therefore, lies in their direct search. While the increasing center-of-mass energy at the LHC can 
probe heavier exotic particles, extracting any new physics information from the tremendous amount of collected 
data also faces the increasing challenge of tackling the QCD background. Hence looking for lepton-enriched 
final states is understandably efficient in suppressing the SM background contributions and probing new 
physics scenarios which can potentially give rise to lepton-rich final states.          

In this work, we consider a 2HDM where the additional Higgs doublet has an odd $\mathcal{Z}_2$ 
symmetry charge opposite to all the SM particles, preventing it from interacting 
directly with the leptons and quarks. One can additionally incorporate right-handed neutrinos 
in the model with similar transformation property under $\mathcal{Z}_2$ symmetry as the new 
Higgs doublet. One can thus generate Dirac neutrino mass terms when the $\mathcal{Z}_2$ 
breaks spontaneously and the new Higgs doublet gets a vacuum expectation value (VEV). This class 
of models, known as neutrinophilic Higgs doublet models ($\nu$HDM) have been proposed long ago 
\cite{Ma:2000cc,Gabriel:2006ns,Davidson:2009ha} and the relevant phenomenology has been studied quite extensively 
\cite{Gabriel:2008es,Haba:2011nb,Haba:2011fn,Chao:2012pt,Maitra:2014qea,Chakdar:2014ifa, Seto:2015rma, Bertuzzo:2015ada, 
Guo:2017ybk}. In principle, one can also generate Majorana neutrino mass terms in such a scenario, 
since a Majorana mass term for the additional right-handed neutrinos does not break the 
$\mathcal{Z}_2$ symmetry but breaks the accidental lepton number symmetry by two units ($\Delta L=2$). 
Such neutrino mass generation mechanism looks very similar to the Type-I seesaw 
\cite{Minkowski:1977sc, Mohapatra:1979ia, Yanagida:1979as,GellMann:1980vs} case, save for the fact 
that one uses the neutrinophilic Higgs VEV instead of electroweak VEV in order to generate the 
light-heavy neutrino mixing. The advantage of having the additional Higgs doublet to generate 
non-zero neutrino masses is that the additional VEV can be very small\footnote{This
is also preferred from naturalness argument \cite{tHooft:1979rat}.} in order to counter the 
smallness of the light neutrino masses which would otherwise be fit with a 
very small Dirac neutrino Yukawa coupling that has no significant collider phenomenological aspects. 

Depending on whether the non-zero neutrinos are Dirac or Majorana type, the collider signals of 
a $\nu$HDM scenario can be very different. When Majorana neutrinos exist, smoking gun 
signal would be lepton number violating final states. In this work, instead of looking for direct 
heavy neutrino production, we have considered the production of the neutrinophilic charged Higgs 
($H^{\pm}$) and explored its various possible decay modes. There are some earlier studies on 
the charged Higgs in similar scenarios emphasising its decay into a charged lepton and a heavy 
neutrino in the process \cite{Guo:2017ybk}. We show that even cleaner signals can be obtained 
using this decay mode with higher lepton multiplicity where the SM background is practically 
non-existent. We also show that sizable signal event rates can be obtained with other 
possible decay modes of the $H^{\pm}$, which can serve as complementary 
channels in probing a $\nu$HDM-like scenario. We perform our analysis using the
13 TeV LHC as well as an $e^+e^-$ collider with 1 TeV center-of-mass energy. In the process, 
one can extract information on 
the neutrino sector parameters also. We show that a very clean indication of the neutrino 
mass hierarchy can be obtained from the multiplicity of the charged leptons in the final state 
even after a rigorous collider simulation. Such information can be very useful in 
complementing the neutrino oscillation experiments. 
\section{Model}
\label{sec:model}
In the $\nu$HDM model, the particle content of the SM is extended by one additional Higgs doublet 
($\phi_{\nu}$) and three generations of SM gauge singlet right-handed neutrinos ($N$). A 
discrete ${\mathcal Z}_2$ symmetry is introduced, under which both $\phi_{\nu}$ and $N_i$, $i=1,2,3$, 
are odd while all the SM fields are even. The most general scalar potential involving the two 
Higgs doublets is given by 
\begin{eqnarray}
V_{\rm sc}=-m_1^2\phi^{\dagger}\phi + m_2^2\phi_{\nu}^{\dagger}\phi_{\nu} - 
m_3^2 (\phi^{\dagger}\phi_{\nu} + \phi_{\nu}^{\dagger}\phi) + \frac{\lambda_1}{2}(\phi^{\dagger}\phi)^2 + 
\frac{\lambda_2}{2}(\phi_{\nu}^{\dagger}\phi_{\nu})^2 \nonumber \\ 
+ \lambda_3(\phi^{\dagger}\phi)(\phi_{\nu}^{\dagger}\phi_{\nu}) + 
\lambda_4(\phi^{\dagger}\phi_{\nu})(\phi_{\nu}^{\dagger}\phi) +
\frac{\lambda_5}{2}[(\phi^{\dagger}\phi_{\nu})^2 + (\phi_{\nu}^{\dagger}\phi)^2],
\label{eq:sc_pot}
\end{eqnarray}
where a non-zero $m_3$ explicitly breaks the ${\mathcal Z}_2$ symmetry in the model. In the absence 
of this term, the ${\mathcal Z}_2$ symmetry can be broken spontaneously by a vacuum 
expectation value (VEV) $v_{\nu}$ of the field $\phi_{\nu}$, while the standard electroweak 
symmetry is broken when $\phi$ acquires a VEV, $v_{\phi}$.

Let us first discuss a framework, where $m_3=0$, {\it i.e.} ${\mathcal Z}_2$ symmetry is broken only spontaneously 
in order to generate light neutrino masses and mixing.
The model is constrained by sterile neutrino searches, effective number of neutrinos and amount of $^4$He 
required in big bang nucleosynthesis (BBN), observed temperature anisotropies of cosmic microwave background (CMB) and astrophysical limits.

Due to an instability of right-handed neutrinos $N_i$ induced by their mixing to left-handed neutrinos, the mixing strength between 
$\nu_\ell$, $\ell =e,\mu,\tau$, and $N_i$, that is, $|U_{\ell i}|$, can be probed by sterile neutrino searches. In semileptonic meson 
decays, $N_i$ are produced, and can subsequently decay to charged leptons and mesons. Present constraints on $|U_{ei}|^2$ and $|U_{\mu i}|^2$ 
allow a region where their magnitude is of order $10^{-10}$ to $10^{-6}$, assuming $M_{N_i} < 2$ GeV \cite{Deppisch:2015qwa}. 
For tau-sterile mixing, $|U_{\tau i}|^2 \lesssim 10^{-4}$, assuming $M_{N_i} < 0.3$ GeV. 

\begin{figure}
\begin{center}
\includegraphics[scale=0.4]{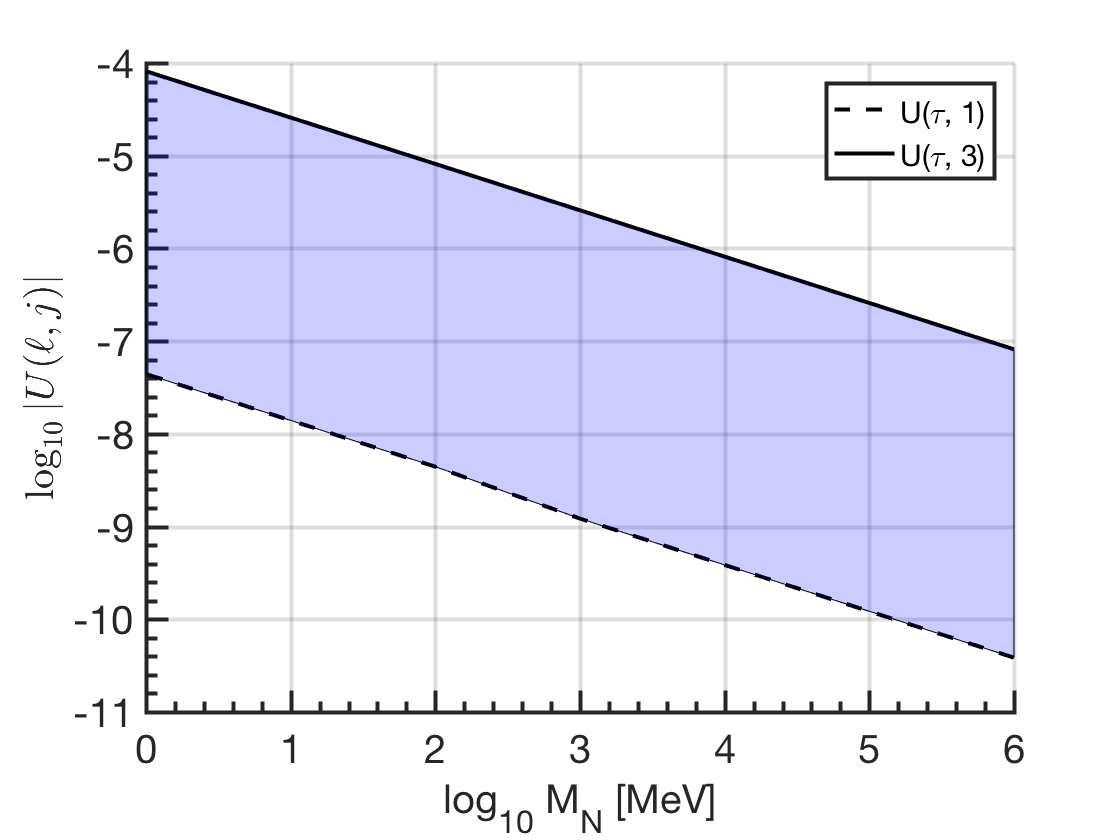}
\end{center}
\caption{\label{Fig:LR}Absolute values of active-sterile mixing block matrix elements as a function of heavy neutrino mass $M_N$. All the 
active-sterile elements fall in the blue band.}
\end{figure}

In $\nu$HDM, however, we found the model favoring even lower values 
of active-sterile mixing, of order $|U_{\ell i}|^2 \sim 10^{-18}$ to $10^{-12}$, at $M_{N_i} = 1$~GeV, and even lower for higher Majorana 
neutrino masses (see Fig. \ref{Fig:LR}).
The largest and smallest active-sterile mixings are driven by $U_{\tau 3}$ and $U_{\tau 1}$ elements. 
Therefore all the active-sterile mixing elements fall between them: $|U_{\tau 1}| < |U_{\ell i}| < |U_{\tau 3}|$. 
The matrix elements are proportional to $M_N^{-1/2}$, therefore $C_1 \frac{1}{\sqrt{M_N}} < |U_{\ell i}|  < C_2 \frac{1}{\sqrt{M_N}}$
with some constants $C_1$ and $C_2$. They are deduced from Fig. \ref{Fig:LR}, having values $C_1 = 4.34\cdot 10^{-9}$ MeV$^{1/2}$ and 
$C_2 = 2.34\cdot 10^{-6}$ MeV$^{1/2}$. The matrix elements then belong to the following interval:
\begin{equation}
1.4\cdot 10^{-10} \sqrt{\frac{\text{GeV}}{M_{N_i}}} \lesssim |U_{\ell i}| \lesssim 7.4\cdot 10^{-8}\sqrt{\frac{\text{GeV}}{M_{N_i}}}
\end{equation}
In addition, the constraints for $|U_{\ell i}|$ were derived from assumption that the branching ratios for $N_i$ decay are dominant. 
This is not applicable for $\nu$HDM, since then the decay modes of right-handed neutrinos are dominated by decays to invisible particles.

As the model is unconstrained by semileptonic and leptonic decay modes, the lower bound for $M_{N_i}$ arises from BBN. 
In the early universe the right-handed neutrinos must be heavy enough to fall off from the thermal equilibrium before BBN. 
This is due to the latest results for effective number of neutrinos ($N_\nu = 3.15 \pm 0.23$) by PLANCK \cite{Planck}, 
which forbids large interference from right-handed neutrinos. This leads to a constraint $M_{N_i} \gtrsim 100$ MeV.

In addition neutrinophilic VEV $v_\nu$ is constrained from both above and below. Ultralight VEV is forbidden by astrophysical 
constraints: $v_\nu \gtrsim \mathcal{O}$(eV) \cite{Zhou,Sher}. On the other hand, the surface energy density associated with the 
domain wall arising from discrete $\mathcal{Z}_2$ symmetry breaking is $\eta \sim v_\nu^3$ \cite{Zeldovich:1974uw}. The effect of 
these domain walls to the temperature anisotropies of CMB is	
\begin{equation}
\frac{\Delta T}{T} \approx \frac{G\eta}{H_0}
\end{equation}
where $G$ is Newton's gravitational constant, $H_0$ is Hubble constant and we have assumed $\lambda_i < 1$ \cite{Yao}. Since the 
observed temperature anisotropies by PLANCK are $\sim 10^{-5}$, the birth of a domain wall will not contradict cosmological data 
if the VEV is small. If we require the contribution to CMB temperature anisotropies not to exceed the experimental limit, 
together with the astrophysical constraints, we get
\begin{equation}
\mathcal{O}(\mathrm{eV}) \lesssim v_\nu \lesssim \mathcal{O}(\mathrm{MeV})
\end{equation}
\begin{figure}
\begin{center}
\includegraphics[scale=0.4]{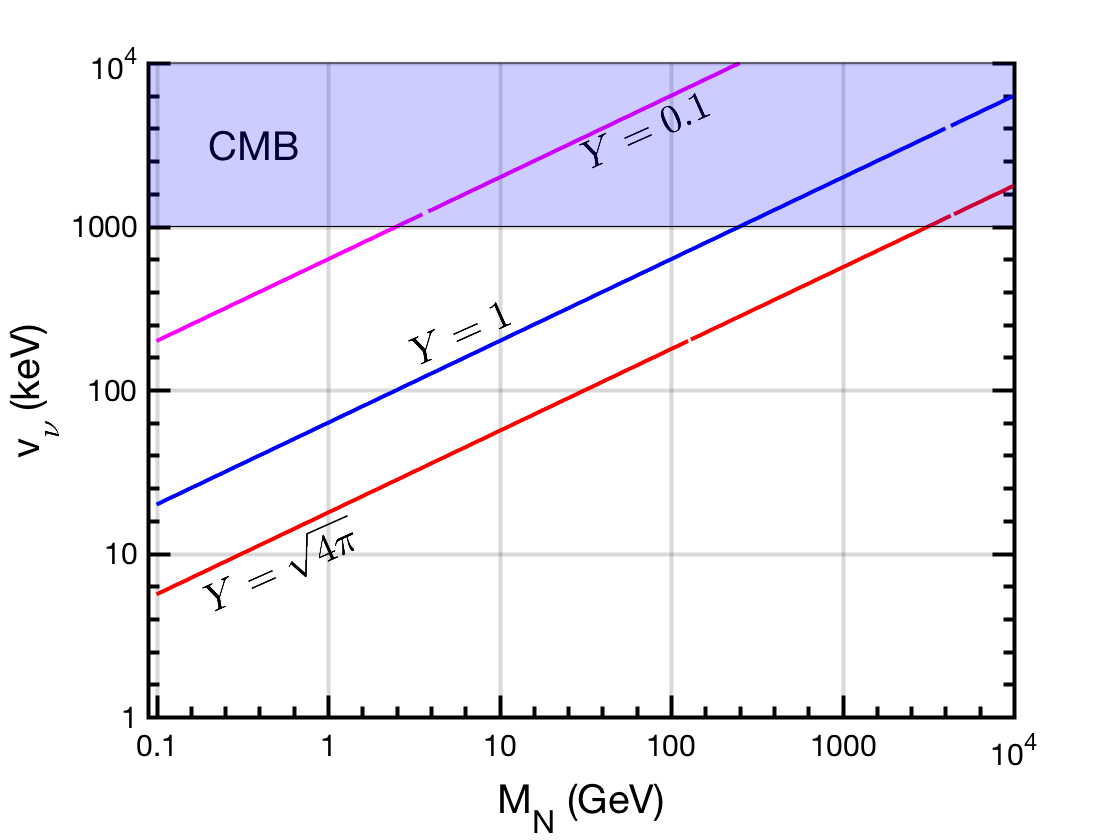}
\end{center}
\caption{\label{Fig:Y}Yukawa contours on $(M_{N_i},v_\nu)$ plane. The lines corresponding to the neutrino Yukawa couplings 
$Y=0.1,1,\sqrt{4\pi}$ are drawn. Below the red $Y=\sqrt{4\pi}$ line, the theory is nonperturbative. Blue-shaded region denoted 
'CMB' is excluded due to restrictions of CMB temperature anisotropies induced by domain walls. The available parameter space is 
restricted also from BBN requirement $M_{N_i} \gtrsim 0.1$ GeV.}
\end{figure}
%
In order to apply perturbative theory to $\nu$HDM, the absolute values of the elements of the light neutrino Yukawa coupling 
matrices must be $\sqrt{4\pi}$ at most. We performed a global fit to available neutrino oscillation data to calculate the matrix 
elements, assuming normal neutrino mass ordering, higher $\theta_{23}$ octant and no CP violation. We found the dependence of the 
largest Yukawa coupling of $v_\nu$ and $M_N$ to be
\begin{equation}
\max |Y(v_\nu,M_N)| \approx 0.629\times \frac{100\:\mathrm{keV}}{v_\nu}\times \sqrt{\frac{M_N}{100\:\mathrm{GeV}}}
\end{equation}
The dependence is illustrated in Fig.~\ref{Fig:Y}.


The breaking of ${\mathcal Z}_2$ 
symmetry is necessary in order to generate light neutrino masses within the framework of this 
model by means of their mixing with heavy right-handed neutrinos. One can add the following 
Yukawa interaction and Majorana neutrino mass terms to the Lagrangian while keeping the 
${\mathcal Z}_2$ parity unbroken:  
\begin{eqnarray}
{\cal L}_{\rm add} = y_{\nu}^{ij}\bar L_i\phi_{\nu}N_j + \frac{1}{2} m_R^{ij} N_i^c N_j + h.c. 
\end{eqnarray} 
where, $m_R^{ij}$ represents the Majorana mass terms corresponding to the right-handed neutrinos. 
Once $\phi_{\nu}$ acquires a VEV, the Yukawa term gives rise to Dirac neutrino mass terms, $m_D^{ij}=v_{\nu}\times y_{\nu}^{ij}$.  

The physical Higgs sector now consists of two neutral CP-even ($h$, $\hh$), one neutral CP-odd ($\ah$) 
and the charged Higgs ($\hh^\pm$)\footnote{We have assumed the scalar potential to be CP invariant.}. 
In the case when $m_3= 0$, the physical mass eigenvalues at tree 
level are given by:
\begin{eqnarray}
m_{h}=\sqrt{\lambda_1 v_{\phi}^2}, ~\mhh=\sqrt{\lambda_2v_{\nu}^2}, ~\mah=\sqrt{-\lambda_5v^2}, 
~\mch=\sqrt{-\frac{v^2}{2}(\lambda_4 + \lambda_5)}
\label{eq:sc_mass_nom3}
\end{eqnarray}
where, $v=\sqrt{v_{\phi}^2 + v_{\nu}^2}$. $v_{\nu}$ being small, terms proportional to 
$v_{\nu}^n ({\rm where} ~n> 2)$ have been neglected. Note that the mixing angle between 
the SM and neutrinophilic Higgs states are proportional to the ratio $\frac{v_{\nu}}{v_{\phi}}$ 
and can be safely neglected since we assume $v_{\nu} \ll v_{\phi}$. Under this circumstance, the 
CP-even neutrinophilic Higgs ($\hh$) is always light and the heavy neutrino almost always decay into 
$\hh$ and a light neutrino resulting in an opposite-sign dilepton signal for a charged Higgs pair 
production channel \cite{Maitra:2014qea}. However, if the explicit symmetry breaking term is present 
in the Lagrangian, i,e, $m_3\neq 0$, the mass eigenvalues are given by: 
\begin{eqnarray}
m_{h}=\sqrt{\lambda_1 v_{\phi}^2}, ~\mhh=\sqrt{ m_3^2\frac{v_{\phi}}{v_{\nu}} + \lambda_2v_{\nu}^2}, \nonumber \\ 
~\mah=\sqrt{m_3^2\frac{v_{\phi}}{v_{\nu}}-\lambda_5v^2}, 
~\mch=\sqrt{m_3^2\frac{v_{\phi}}{v_{\nu}} - \frac{v^2}{2}(\lambda_4 + \lambda_5)}
\label{eq:sc_mass_m3}
\end{eqnarray}
Now the neutrinophilic CP-even Higgs can be heavy depending on our choice of $m_3$, 
$\mhh \simeq \sqrt{m_3^2\frac{v_{\phi}}{v_{\nu}}}$. A heavy $\hh$ and (or) 
$A_\nu$ opens up the possibility of a cascade decay via heavy neutrinos resulting 
in multi-lepton signals of such a scenario that we intend to explore. In the limit $m_3\to 0$, 
the symmetry of the theory is enhanced. Thus, $m_3$ can be assumed to be naturally small. Besides, a large $m_3$ 
can also give rise to significant mixing between the two Higgs doublets, which is strictly constrained 
from the present Higgs data. 
\subsection{Neutrino Mass Generation}
\label{sec:nu_mass_gen}
The neutrino oscillation data \cite{GonzalezGarcia:2007ib,Gonzalez-Garcia:2015qrr,deSalas:2017kay} indicates 
that at least two of the three light neutrinos have non-zero mass. 
One of the most natural ways to generate tiny neutrino mass is via seesaw mechanism \cite{Minkowski:1977sc,Mohapatra:1979ia, 
Yanagida:1979as, GellMann:1980vs,Glashow:1979nm,Schechter:1981cv,Schechter:1980gr, Weinberg:1979sa, 
Weinberg:1980bf}. In $\nu$HDM  the mechanism is very similar to that of Type-I 
seesaw \cite{Minkowski:1977sc, Mohapatra:1979ia,Yanagida:1979as,GellMann:1980vs}. The mixing between 
light and heavy neutrinos is introduced via the term $y_{\nu} \bar L \phi_{\nu} N$ in the aftermath 
of symmetry breaking, when $\phi_{\nu}$ gets a VEV. In the basis $\lbrace\nu, N\rbrace$ the $6\times 6$ 
neutrino mass matrix looks like
\begin{eqnarray}
{\cal M}_{6\times 6}=\left(\begin{array}{cc}
{0}_{3\times 3} & {m_D}_{3\times 3}\\
{m_D^T}_{3\times 3} & {m_R}_{3\times 3}
\end{array}\right),
\label{eq:nmfull}
\end{eqnarray}
where $m_D=y_{\nu}v_{\nu}$. The light effective $3\times 3$ neutrino mass matrix in the approximation 
$m_D << m_R$ is given by
\begin{eqnarray}
M_{\nu} = m_D m_R^{-1} m_D^T .
\label{eq:nmlight}
\end{eqnarray} 
The above equation looks exactly similar to what we obtain in canonical Type-I seesaw scenario. 
The only difference is that in the present framework $v_{\nu}$ can be quite small and as a result 
one can have larger $y_{\nu}$ compared to the canonical Type-I seesaw scenario, thus making 
this model phenomenologically more interesting. In order to fit the oscillation data, one also 
needs to account for the mixing among the three light neutrino states constrained by the PMNS 
matrix. One can rewrite $M_{\nu}$ in Eq.~\eqref{eq:nmlight} as 
\begin{eqnarray}
M_{\nu} = U^T m_{\nu}^{\rm diag}U,
\label{eq:nmlight_ca1}
\end{eqnarray} 
where $m_{\nu}^{\rm diag}$ is the diagonal light $3\times 3$ neutrino mass matrix and $U$ is the 
PMNS mixing matrix. In order to produce proper mixing satisfying the experimental bounds on the PMNS 
matrix elements, one of the matrices, $m_D$ or $m_R$, has to be off-diagonal. Here we 
choose to keep $m_R$ diagonal and fit the PMNS matrix via an off-diagonal $m_D$. Thus $y_{\nu}$ is 
obtained using Casas-Ibarra parameterization \cite{Casas:2001sr}
\begin{eqnarray}
y_{\nu} =\frac{1}{v_{\nu}} \sqrt{m_R^{\rm diag}} R \sqrt{m_{\nu}^{\rm diag}} U^T,
\label{eq:nmlight_ca2}
\end{eqnarray}
where $R$ can be any orthogonal matrix and complex provided $R^TR=1$. For simplicity, we have 
chosen $R$ to be an identity matrix. 

Thus  with correct choices of the parameters $m_D$ and $m_R$, Eq.~\eqref{eq:nmfull} is capable of 
explaining the neutrino oscillation data at the tree level itself. There is a potential 
source of large correction \cite{Pilaftsis:1991ug,Grimus:2002nk} to the neutrino states at one 
loop arising from the $H_{\nu} (A_{\nu})$ loops. These mass corrections can be sizeable enough to 
violate the experimental limits. However, the loop contributions to the neutrino masses corresponding 
to $H_{\nu}$ and $A_{\nu}$ have a mutual sign difference and can exactly cancel each other if they are mass 
degenerate \cite{AristizabalSierra:2011mn, AristizabalSierra:2011gg, AristizabalSierra:2011aa,Haba:2011nb}. 
As can be seen both from Eq.~\eqref{eq:sc_mass_nom3} and \eqref{eq:sc_mass_m3}, the mass splitting between 
these two states is driven by the parameter $\lambda_5$ which is therefore set equal to zero throughout 
this work. 
\section{Constraints and Benchmark Points}
\label{sec:cons_bp}
Constraints on the charged Higgs mass and its couplings may arise from direct collider search 
results, neutrino oscillation data and lepton flavor violating decay branching ratios. The 
LHC collaborations have looked for signatures of exotic scalars in various channels and put 
bounds on the charged Higgs mass in the range 300 - 1000 GeV provided it can 
decay only into a top and a bottom quark \cite{ATLAS:2016qiq,Mccarn:2016fuu,Ohman:201637, 
Akeroyd:2016ymd}. However, in our present scenario, the charged Higgs, being a neutrinophilic 
one, does not couple to the quarks. In such scenarios, there are no direct search constraints 
on $\mch$. 
In principle, the constraints derived from slepton searches at the LHC can be reinterpreted 
to put bounds on the neutrinophilic charged Higgs masses although only in the massless limit of the 
lightest neutralino. Two body decay of the sleptons into a charged lepton and lightest neutralino 
gives rise to a dilepton signal which can be relevant for the the present scenario.
Existing data excludes slepton masses upto 450 GeV in presence of massless neutralino \cite{Aad:2014vma,CMS:2017mkt}.
However, one always obtains same-flavor-opposite-sign(SFOS) lepton pairs from such slepton pair production processes. 
The signal requirement also demands a jet veto in the central region alongside the SFOS lepton pair 
for such analyses. In the present scenario, largest event rate in such a signal region can be obtained 
when $\ch$ decays into a charged lepton and a heavy neutrino. Heavy neutrino further decays into a light neutrino 
and Z-boson which further decays invisibly. Clearly, the resulting signal cross-section is rendered small due 
to branching suppressions. Demand of SFOS lepton pairs makes this cross-section even smaller\footnote{The obtained 
signal cross-section for our lightest benchmark point even before the detector simulation is less than the observed 
number as quoted in \cite{Aad:2014vma,CMS:2017mkt}.}. Thus, the 
existing slepton mass limit when reinterpreted for $\mch$ proves to be much weaker.  
Its couplings with the heavy neutrinos on the other hand, can be constrained 
from neutrino oscillation data and lepton flavor violating decay branching ratios \cite{Guo:2017ybk}. As mentioned in 
section~\ref{sec:nu_mass_gen}, we have used off-diagonal $m_D$ while fitting the PMNS matrix. 
These off-diagonal entries are severely constrained from LFV decay branching ratio constraints 
\cite{Adam:2013mnn,Baldini:2013ke,Aubert:2009ag,Aushev:2010bq,Hayasaka:2010np,Bertl:2006up, 
Bartoszek:2014mya}. These constraints are also reflected upon our choice of the neutrinophilic Higgs 
VEV, $v_{\nu}$. It has been observed and also verified by us that $v_{\nu}$ can be $\sim 10^{-2}$ GeV \cite{Guo:2017ybk} at 
the smallest, if the neutrino oscillation data and the LFV constraints are to be satisfied 
simultaneously, the most stringent constraint arising from the non-observation of BR($\mu\to e\gamma$) 
\cite{Adam:2013mnn,Baldini:2013ke}. This constraint puts the spontaneously breaking ${\mathcal Z}_2$ 
scenario in jeopardy. As evident from Fig.~\ref{Fig:Y}, such a choice of $v_{\nu}$ is clearly ruled out 
from restrictions on CMB temperature anisotropies induced by domain walls. However, if the ${\mathcal Z}_2$ 
symmetry is broken explicitly, this domain wall problem can be averted. Hence for this work, we choose to work with 
the $m_3\ne 0$ scenario only. 

\subsection{Charged Higgs branching ratios and pair production cross-section}
\label{sec:chhig}
The possible decay modes of the neutrinophilic charged Higgs ($H_\nu^{\pm}$) in our present scenario 
are $\ch\to\hn\ell^{\pm}$, $\ch\to\hn\tau^{\pm}$, $\ch\to\hh W^{\pm}$ and $\ch\to\ah W^{\pm}$. 
The relevant interaction vertices are given in the Appendix.
Depending on the mass hierarchy of $\ch$, $\hh$($\ah$) and $\hn$, and the choice of neutrino 
mass hierarchy one (or two) of these decay modes determine the event rates of the different possible
final states at the collider. Note that the branching ratios of the decays into the neutral CP-even 
and CP-odd Higgs states are always the same since they are mass degenerate by our choice of the parameters. 
These two decay modes dominate over the heavy neutrino decay modes always, if the mass difference, 
$\Delta m =\mch - \mhh (\mah)$ is larger than that of the W-boson mass, $m_W$. This is an artefact of 
the small Dirac neutrino Yukawa parameters, which are otherwise constrained by neutrino oscillation data 
and the non-observation of LFV decays. The $y_{\nu}$ being smaller by orders of magnitude from the competitive 
gauge coupling, a large branching ratio into the $\hn\ell^{\pm}$ or $\hn\tau^{\pm}$ decay modes are not ensured 
even if $\Delta m < m_W$. In spite of the additional phase space suppression, three-body decays of $\ch$ via 
off-shell $W$-decay, dominate over these two-body modes unless $\Delta m \ll m_W$. 
\begin{figure}[h!]
\begin{center}
\includegraphics[scale=0.45]{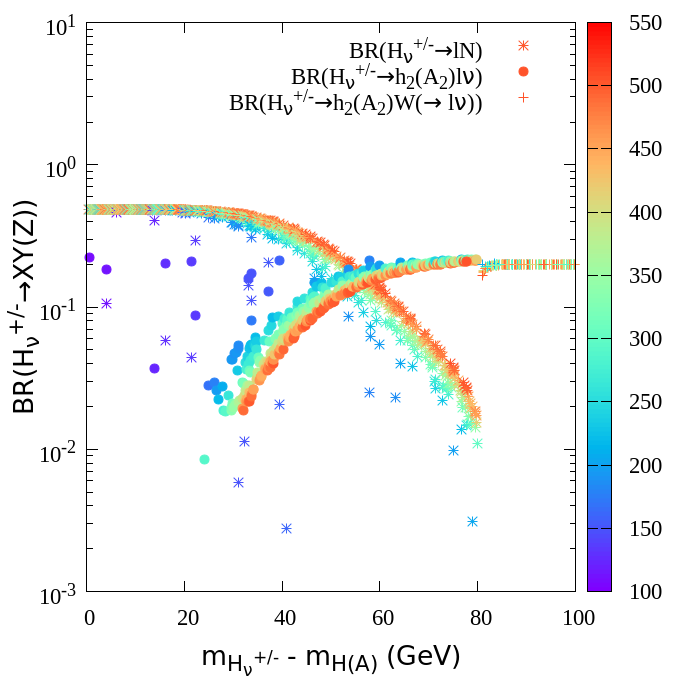} 
\caption{Variation of BR($\ch\to\hn\ell^{\pm}$), BR($\ch\to\hh (\ah) W^{\pm}(\to\ell^{\pm}\nu)$) and BR($\ch\to\hh(\ah)\ell\nu$) as 
a function of $\Delta m$. The color coded bar on the right shows the variation of $\mch$. For all the points, $\mhn$ 
is kept fixed at 100 GeV. Normal hierarchy is assumed for the light neutrinos. For inverted hierarchy, although the numerical values of the 
BRs are expected to be different, the pattern of the distribution remains same.}
\label{fig:ch_br}
\end{center}
\end{figure}
This behavior is depicted in Fig.~\ref{fig:ch_br} where the competitive nature of BR($\ch\to\hn\ell^{\pm}$) 
and BR($\ch\to\hh(\ah)\ell\nu$), where $\ell=e, \mu$, is clearly visible through the distributions of the starred and circular points 
respectively.  BR($\ch\to\hn\ell^{\pm}$) overtakes the three-body decay branching ratio only if $\Delta m < 50$ GeV. 
For $\Delta m > m_W$, BR($\ch\to\hh (\ah) W^{\pm}(\to\ell^{\pm}\nu)$) takes over and remains the only 
dominant decay mode.

\begin{figure}[h!]
\begin{center}
\includegraphics[height=6.5cm,width=7.5cm]{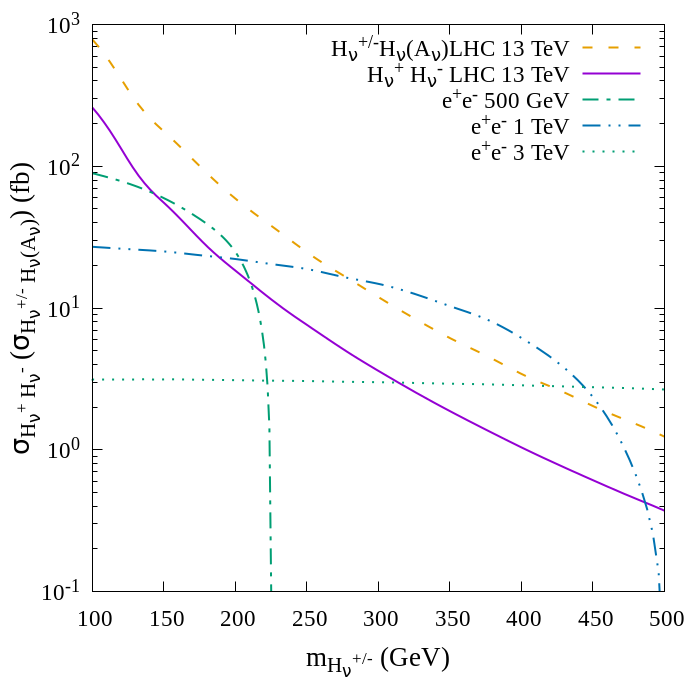} 
\includegraphics[height=6.5cm,width=7.5cm]{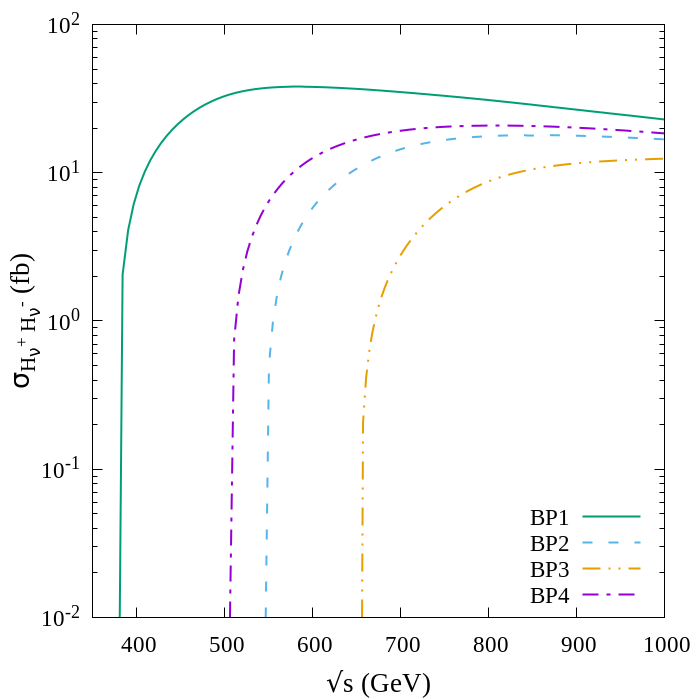} 
\caption{Variation of the charged Higgs pair production cross-section at the LHC and an $e^+e^-$ collider at center-of-mass 
energies of 13 TeV and 1 TeV, respectively. The distribution on the right shows variation of the charged Higgs pair production cross-section at 
an $e^+e^-$ collider with varying center-of-mass energy ($\sqrt{s}$) for our four benchmark points.}
\label{fig:ch_cs}
\end{center}
\end{figure}
In Fig.~\ref{fig:ch_cs}, we have shown variation of the $\ch$ production cross-section at the LHC and an $e^+e^-$ collider.
The figure on the left shows the variation of the cross-sections as a function of $\mch$ at the 13 TeV LHC and different center-of-mass energies 
(500 GeV, 1 TeV and 3 TeV) at an $e^+e^-$ collider. Note that, at the LHC, the $\ch$ production channels include $pp\to H_{\nu}^{\pm}H_{\nu}^{\mp}$, 
$pp\to\ch\hh$ and $pp\to\ch\ah$ while for the $e^+e^-$ collider, pair production is the only viable option. Since we have assumed $\mhh=\mah$ for our 
study, the cross-sections of the above mentioned second and third production channels exactly equal. Hence we have shown their combined cross-section 
in the figure and evidently, it dominates over the pair production cross-section throughout the entire charged Higgs mass range.  
However, both these cross-sections fall rapidly with increasing mass. On the other hand, at an 
$e^+e^-$ collider the cross-section falls far less rapidly implying the fact that such a collider will be more effective than the LHC 
in order to probe heavier charged Higgs masses. The figure on the right shows the variation of the pair production cross-section at an $e^+e^-$ collider 
with varying center-of-mass energies for our chosen benchmark points. Moreover, a lepton collider is likely to be much cleaner in terms of the SM background contributions. 
In this work, we have taken into account all the aforementioned production channels for LHC and just the pair production for the $e^+e^-$ collider analysis. 
\subsection{Choice of benchmark points}
\label{sec:bps}
We now proceed to choose some benchmark points representing the different interesting features of the present 
scenario for further collider studies. 
As discussed earlier, one can obtain different possible final states depending upon the mass hierarchies of $\ch$, $\hh$ ($\ah$) and $\hn$. 
Since we also aim to correlate the light neutrino mass hierarchy with the multiplicity of different lepton flavor final states, 
we will study cases in which 
at least one of the heavy neutrinos is lighter than the neutrinophilic Higgs states so that it can appear in the cascade.  
In Table~\ref{tab:bp} below we present the input parameters, relevant masses and the resulting $y_{\nu}$ for the four 
benchmark points of our choice. We have incorporated the complete model in SARAH \cite{Staub:2008uz,Staub:2009bi,
Staub:2010jh,Staub:2013tta,Staub:2015kfa}, and subsequently imported in SPheno \cite{Porod:2003um,Porod:2011nf} in order 
to perform the analytical and numerical computation of the masses and mixings of the particles, their branching 
ratios and other relevant constraints.
See Appendix for LFV constraints for our benchmarks.
\begin{table}[ht]
\begin{center}
\begin{tabular}{||c|c|c|c|c||} 
\hline\hline 
Parameters & {\bf BP1} & {\bf BP2} & {\bf BP3} & {\bf BP4}  \\
\hline\hline 
$\lambda_1$ &0.270 &0.210 &0.235 &0.212 \\
$\lambda_2$ &0.50 &0.50 &0.50 &0.50 \\
$\lambda_3$ & 1.50& 1.50& 1.50 & 1.50\\
$\lambda_4$ & -0.01& -1.50& -0.01& -1.10\\
$m_3^2~{\rm GeV}^2$  & -1.50& -1.50& -4.50& -1.50\\
$m_R^{ii}$ (GeV) & 100.0& 100.0& 200.0 & 125.0\\
\hline\hline
$\mhh,\mah$ (GeV) & 187.5& 187.9& 325.6 & 188.5\\
$\mch$ (GeV) &  188.5& 272.8& 326.4 & 252.8\\
$m_N$ (GeV) &  100.0&  100.0& 200.0 & 125.0\\
\hline
 & (Normal) & (Normal) & (Normal) & (Normal)\\
$y_{\nu}$ & {\tiny $\left(\begin{array}{ccc}
1.445 & 2.261  & 0.336 \\
2.261& 5.719 & 3.396\\
0.336 & 3.396 & 6.944\\
\end{array}\right)$} & {\tiny $\left(\begin{array}{ccc}
1.445 & 2.261  & 0.336 \\
2.261& 5.719 & 3.396\\
0.336 & 3.396 & 6.944\\
\end{array}\right)$} & {\tiny $\left(\begin{array}{ccc}
2.044 & 3.197 & 0.476 \\
3.197 & 8.088 & 4.802 \\
0.476 & 4.802 & 9.820\\ 
\end{array}\right)$} & {\tiny $\left(\begin{array}{ccc}
1.616 & 2.528 & 0.376 \\
2.528 & 6.394 & 3.797 \\
0.376 & 3.797 & 7.763 \\
\end{array}\right)$} \\ 
 &(Inverted) & (Inverted) & (Inverted) & (Inverted)\\
($\times 10^{3}$) & {\tiny $\left(\begin{array}{ccc}
3.796 & 7.116 & 5.594\\
7.116 &  0.785&  2.135\\
5.694 & 2.135 & 3.101 \\
\end{array}\right)$} & {\tiny $\left(\begin{array}{ccc}
3.796 & 7.116 & 5.594\\
7.116 &  0.785&  2.135\\
5.694 & 2.135 & 3.101 \\
\end{array}\right)$} & {\tiny $\left(\begin{array}{ccc}
5.369 & 10.060 & 8.053\\
10.060 & 1.109 & 3.019\\
8.053 & 3.019 & 4.386 \\
\end{array}\right)$}& {\tiny $\left(\begin{array}{ccc}
4.245 & 7.956 & 6.367\\
7.956 & 0.877 & 2.387\\
6.367 & 2.387 & 3.467\\
\end{array}\right)$}  \\ 
\hline\hline
\end{tabular}
\caption{Relevant model parameters and masses. As mentioned before the parameter $\lambda_5$ is set equal to zero 
throughout this work.}
\label{tab:bp}
\end{center}
\end{table}

The four benchmark points are chosen such that all the dominant decay modes of the neutrinophilic Higgs and the heavy neutrinos are 
highlighted by different mass hierarchies. The relevant branching ratios are shown in Table~\ref{tab:bp_br}. The two most 
dominant decay modes of $\ch$ are $N\ell$, where $\ell=e,\mu,\tau$, and $\hh(\ah)W^{\pm}$. The first decay mode is driven 
by the Dirac neutrino Yukawa couplings, $y_{\nu}$, whereas the second one is driven by gauge couplings. As discussed above, the elements of 
$y_{\nu}$ are already constrained from the neutrino oscillation data as well as from the LFV constraints, and thus are in general weaker than the 
competitive gauge coupling. Hence, if the mass splittings among the neutral and charged neutrinophilic Higgs and the heavy neutrino states
are such that both $N\ell$ and $\hh(\ah)W^{\pm}$ decay modes are kinematically accessible for $\ch$, the gauge boson associated one becomes its only relevant 
decay mode. However, if at least one of the heavy neutrinos is lighter than the $\ch$ and the $\hh(\ah)$ states are almost degenerate 
to it, then the decay via heavy neutrinos becomes important. The latter scenario is highlighted in {\bf BP1} and {\bf BP3} while {\bf BP2}  
represents the former scenario. {\bf BP4}, on the other hand, highlights the situation where the two-body mode $\ell\hn$ competes with 
the three-body decay into $\hh$ (or $\ah$) alongside an off-shell $W$-boson. However, $\Delta m$ being on the larger side, the three-body decay 
dominates as discussed earlier in Section~\ref{sec:chhig}. The heavy neutrinos ($N$) in this scenario can decay either via the SM gauge bosons 
($W^{\pm}, Z$) or the different Higgs 
states. Note that decays of $N$ into $W^{\pm}$, $Z$ and $h$ can only occour through their mixing with the light neutrinos which are suppressed in the 
present scenario. Hence, these decay modes become relevant for $N$ only if the neutrinophilic Higgs states are kinematically inaccessible to it. 
The choice of neutrino mass hierarchy 
clearly reflects in the branching ratios of both $\ch$ and $N$ and is also expected to be reflected in the final 
event rates of the multi-lepton signals we intend to explore.  
\begin{table}[h!]
\begin{center}
\begin{tabular}{||c||c|c||c|c||c|c||c|c||} 
\hline
\multicolumn{1}{||c||}{Branching} &
\multicolumn{2}{|c||}{\bf BP1} &
\multicolumn{2}{|c||}{\bf BP2} &
\multicolumn{2}{|c||}{\bf BP3} &
\multicolumn{2}{|c||}{\bf BP4} \\
\cline{2-9}
Ratio & Normal & Inverted & Normal & Inverted & Normal & Inverted & Normal & Inverted \\
\hline\hline
BR($\ch\to N\ell^{\pm}$) & 0.49  & 0.77  & - & - & 0.49 & 0.77 & 0.05 & 0.13\\
BR($\ch\to N\tau^{\pm}$) & 0.51  & 0.23  & - & - & 0.51 &0.23 & 0.06& 0.04\\
BR($\ch\to \hh W^{\pm}$) & -  & -  & 0.50 & 0.50 & - & - & -& -\\
BR($\ch\to \ah W^{\pm}$) & -  & -  & 0.50 & 0.50 & - & - & -& -\\
BR($\ch\to \hh\ell\nu$) & -  & -  & - & - & - & - & 0.10& 0.09\\
BR($\ch\to \ah\ell\nu$) & -  & -  & - & - & - & - & 0.10& 0.09\\
\hline
BR($N\to\ell^+ W^-$) & 0.21 & 0.43  & 0.21 & 0.43 & 0.16 & 0.21 & 0.17& 0.23\\
BR($N\to\ell^- W^+$) & 0.21 & 0.43  & 0.21 & 0.43 & 0.16 & 0.21 & 0.17& 0.23\\
BR($N\to\tau^+ W^-$) & 0.23 & 0.01  & 0.23 & 0.01 & 0.18 & 0.13 & 0.20& 0.14\\
BR($N\to\tau^- W^+$) & 0.23 & 0.01  & 0.23 & 0.01 & 0.18 & 0.13 & 0.20& 0.14\\
BR($N\to\nu_{\ell} Z$) & 0.12 & 0.12  & 0.12 & 0.12 & 0.32& 0.32& 0.27& 0.27\\
\hline
BR($\hh (\ah)\to\nu_{\ell}N$) & 1.00 & 1.00  & 1.00 & 1.00 & 1.00 & 1.00& 1.00& 1.00\\
\hline\hline
\end{tabular}
\caption{Relevant branching ratios. Here $\ell =e, \mu$.}
\label{tab:bp_br}
\end{center}
\end{table} 

\begin{figure}
\begin{center}
\includegraphics[scale=0.4]{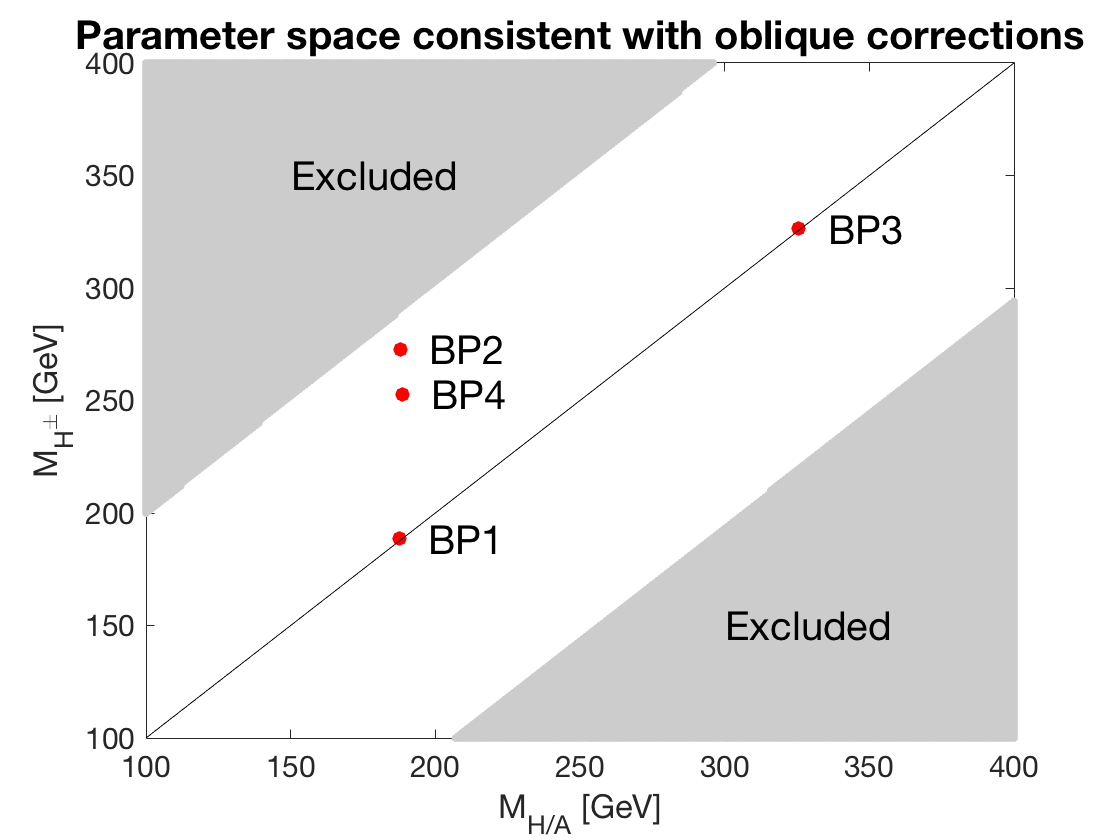}
\end{center}
\caption{\label{Fig:Obl} In the gray region, the oblique corrections induced by the neutrinophilic Higgses are too large. A too large mass difference $\Delta m \equiv  m_{H_\nu^\pm} - m_{H_\nu, A_\nu}$ is disfavored. Red dots label the chosen benchmark points. Black line corresponds to $m_{H_\nu^\pm} = m_{H_\nu, A_\nu}$.}
\end{figure}

In addition, we checked the effect of neutrinophilic Higgses on the oblique parameters $(S,T,U)$. We have ensured the corrections induced by our benchmark points do not exceed the uncertainties given in \cite{He2001}. See Fig. \ref{Fig:Obl} for the allowed $(m_{H_\nu, A_\nu},m_{H_\nu^\pm})$ values.
\section{Collider Analysis}
\label{sec:coll_ann}
Charged Higgs in our benchmark scenarios can give rise to novel signatures in lepton enriched final states. Majorana neutrinos, 
if produced via cascade from the charged Higgs can further decay resulting in same-sign leptonic final states, which 
are characteristic to seesaw models and also have much less SM background. The gauge bosons resulting from the decays 
of the neutrinophilic Higgs and heavy neutrinos may also decay leptonically and thus one can easily obtain a multi-lepton 
final state associated with missing energy. Leptonic branching ratios of the gauge bosons being small, one would expect 
smaller event rates in the final state with increasing lepton multiplicity. However, it also means less SM background to deal 
with resulting in cleaner signals. In this section we explore the different possible multi-lepton final states with or without 
the presence of additional jets along with detailed signal to background simulation in order to ascertain the discovery potential of the 
charged Higgs for our chosen benchmark points in the context of 13 TeV LHC as well as future lepton colliders. 
\subsection{Identifying signal regions}
\label{sec:id_sr}
In the context of LHC, we aim to study cleaner multi-lepton channels with no tagged jets in the final state.   
The possible final states that we probe in the present context are $\ge 6\ell + \met + X$ ({\bf SR1}), 
$\ge 5\ell + \met + X$ ({\bf SR2}), and same-sign trilepton ($SS3\ell$) + $\met + X$ ({\bf SR3}), where  $X$ 
represents everything else (jets, photons or leptons) \footnote{For {\bf SR3}, $X$ consists of no leptons since 
in this case, we demand exactly three leptons with same sign in the final state.} in the final states. As mentioned earlier, 
the various branching ratios of $\ch$ and hence the final signal event rates depend on the mass difference factor $\Delta m$. 
Thus, it is interesting to study how the signal rates vary depending on $\Delta m$ which in turn can also provide an 
indirect hint about the masses of $\hh$ and (or) $\ah$.   
\begin{figure}[h!]
\begin{center}
\includegraphics[height=5.4cm,width=5.4cm]{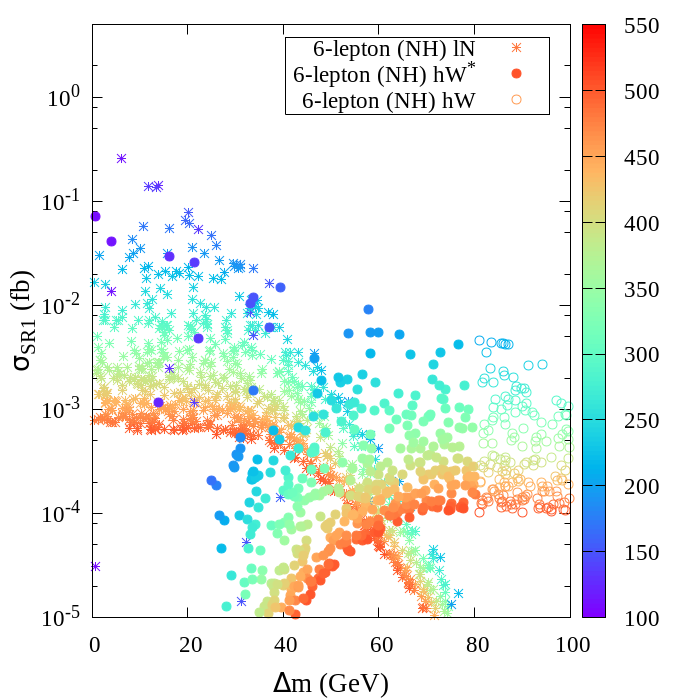} 
\includegraphics[height=5.4cm,width=5.4cm]{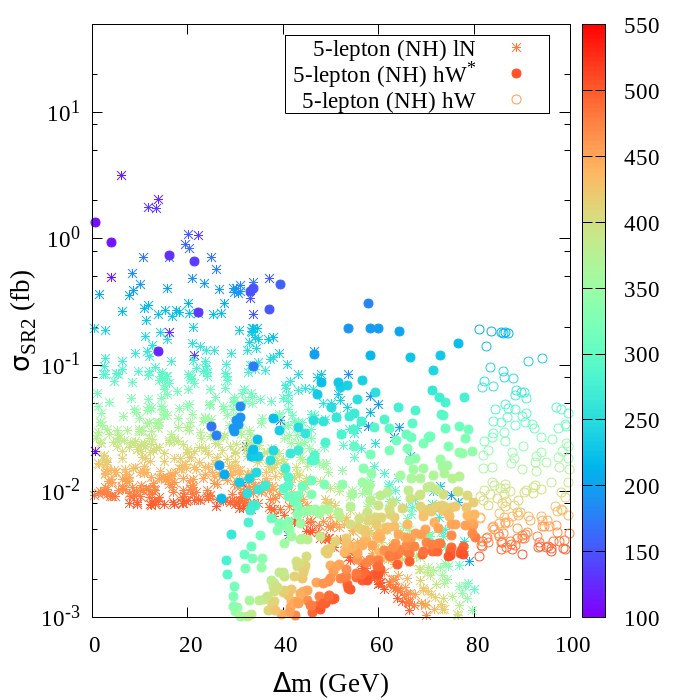} 
\includegraphics[height=5.4cm,width=5.4cm]{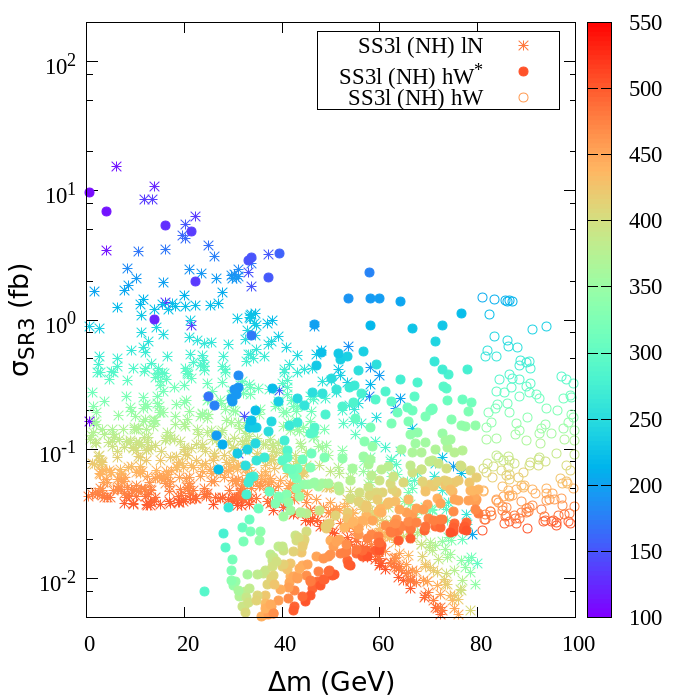} \\ 
\includegraphics[height=5.4cm,width=5.4cm]{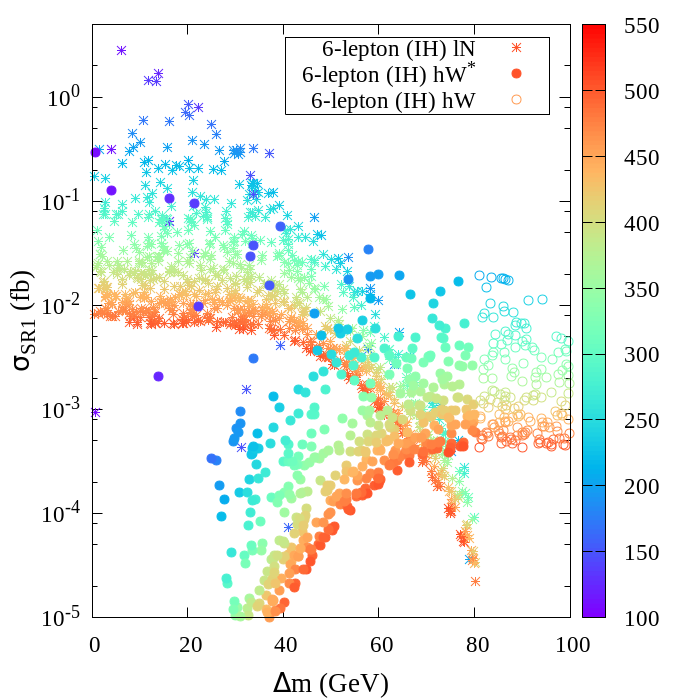} 
\includegraphics[height=5.4cm,width=5.4cm]{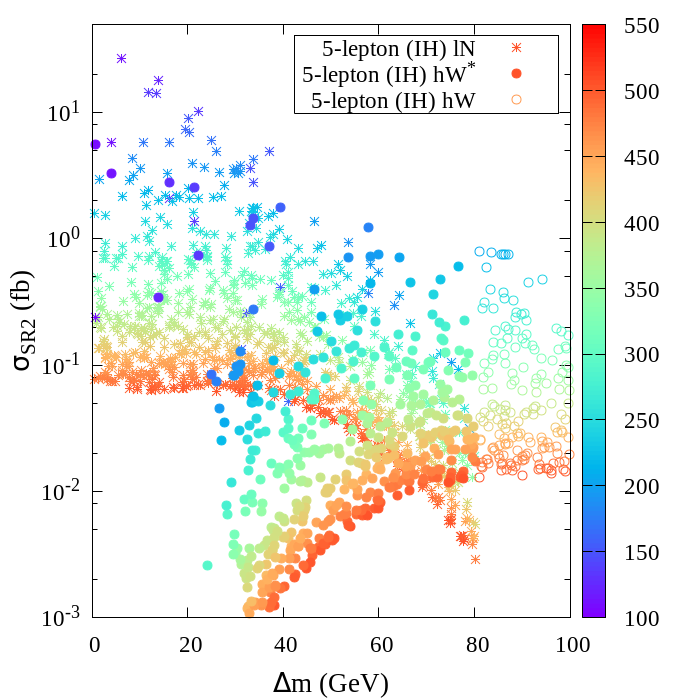} 
\includegraphics[height=5.4cm,width=5.4cm]{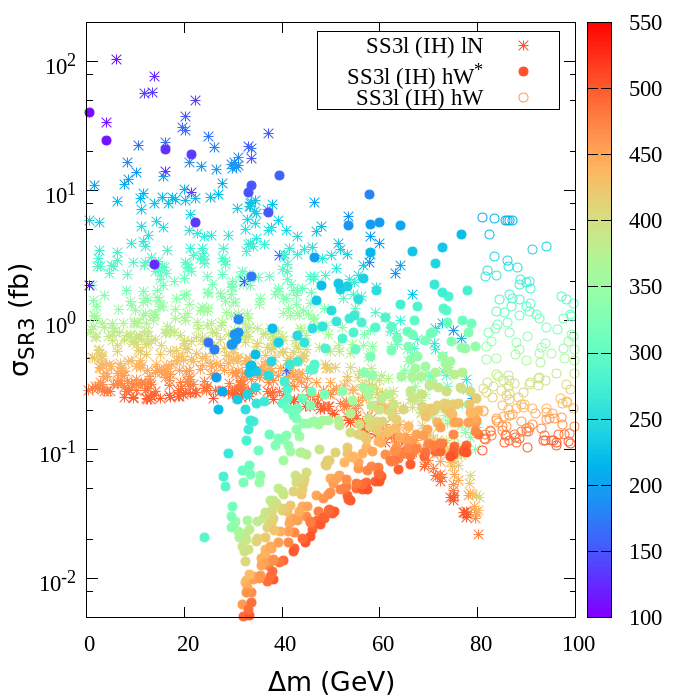} 
\caption{Variation of cross-sections corresponding to the signal regions {\bf SR1}, {\bf SR2} and {\bf SR3} as a function of $\Delta m$ at 13 TeV LHC. 
The two rows represent scenarios with 
normal and inverted hierarchy of the light neutrino masses respectively. The color coding 
represents variation of $\mch$. For all points, heavy neutrino masses are kept at 100 GeV.}
\label{fig:SR_var_lhc}
\end{center}
\end{figure}

In Fig.~\ref{fig:SR_var_lhc} we have shown the variation of the cross-sections corresponding to the three signal regions mentioned above 
as a function of $\Delta m$ with color-coded $\mch$. 
Note that these cross-sections are theoretical estimates obtained after combining contributions from all the three relevant production modes of 
$\ch$ at the LHC prior to detector simulation and do not include the cut efficiencies. The two rows of the figures correspond to normal 
and inverted hierarchy of neutrino masses respectively.
While {\bf SR1} only receives contribution from pair-production, both {\bf SR2} and {\bf SR3} are enriched 
with contributions from pair production as well as associated production of the $\ch$.
Most of the signal events corresponding to {\bf SR1} and {\bf SR2} are expected to arise from $\ch$ decay into a charged 
lepton and a heavy neutrino followed by the heavy neutrino decay into a charged lepton and $W$. Depending on the leptonic or hadronic decays 
of the $W$-bosons, one can obtain various lepton multiplicities as represented by these signal regions. The signal cross-sections are largest 
when $\ch$ and $\hh$ ($\ah$) are mass degenerate for any given $\mch$ and they drop with increasing $\mch$ and $\Delta m$. {\bf SR3}, on the other hand, 
receives more contribution when $\ch$ decays into $\hh$ (or $\ah$) along with an on-shell or off-shell $W$-boson. Moreover, in the case of pair production, 
same-sign leptons can not be obtained if both the $\ch$ decays via $\ell^{\pm}N$ resulting the cross-section  
of {\bf SR3} being smaller with smaller $\Delta m$. However, the associated production channels contribute dominantly to this signal region throughout the whole range 
of $\Delta m$. As evident from Fig.~\ref{fig:SR_var_lhc}, {\bf SR3} is the most favorable channel to look for such scenarios. 
In general, the inverted hierarchy of the light neutrino masses is expected to generate more multi-leptonic events owing to the larger 
branching ratio, BR($\ch\to e\hn$) as reflected by the plots on the bottom row.    

In a similar way, we now proceed to choose some signal regions for our analyses in the context of an $e^+e^-$ collider. The possible final states 
we probe in this context are $\ge 5\ell + \me + X$ ({\bf SR4}), $\ge 4\ell + \ge 2- {\rm jet} + \me + X $ ({\bf SR5}),
and $SS3\ell + \me + X$ ({\bf SR6}) \footnote{Just like {\bf SR3}, $X$ in {\bf SR6} also does not contain any leptons.}.
\begin{figure}[h!]
\begin{center}
\includegraphics[height=5.4cm,width=5.4cm]{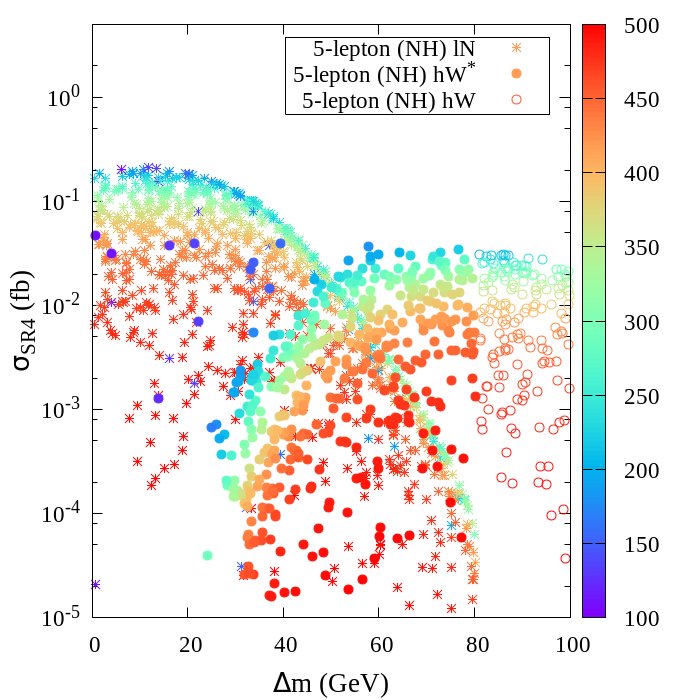} 
\includegraphics[height=5.4cm,width=5.4cm]{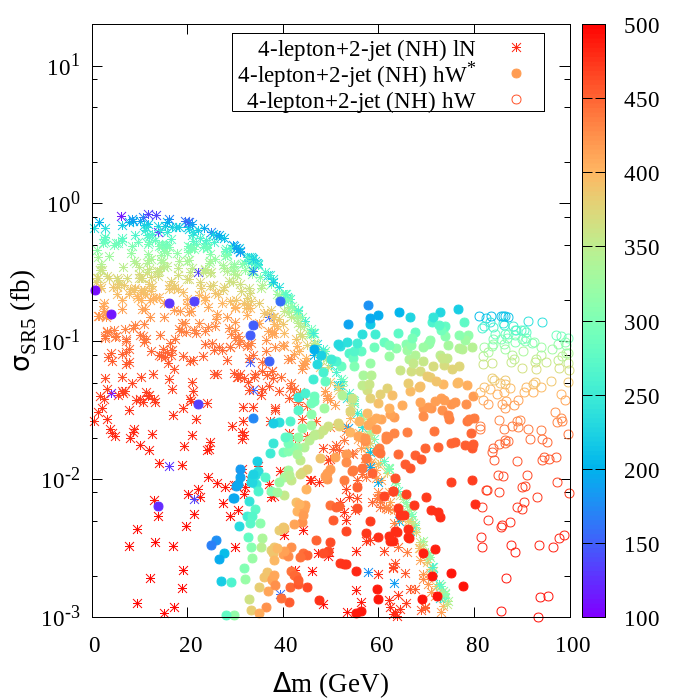} 
\includegraphics[height=5.4cm,width=5.4cm]{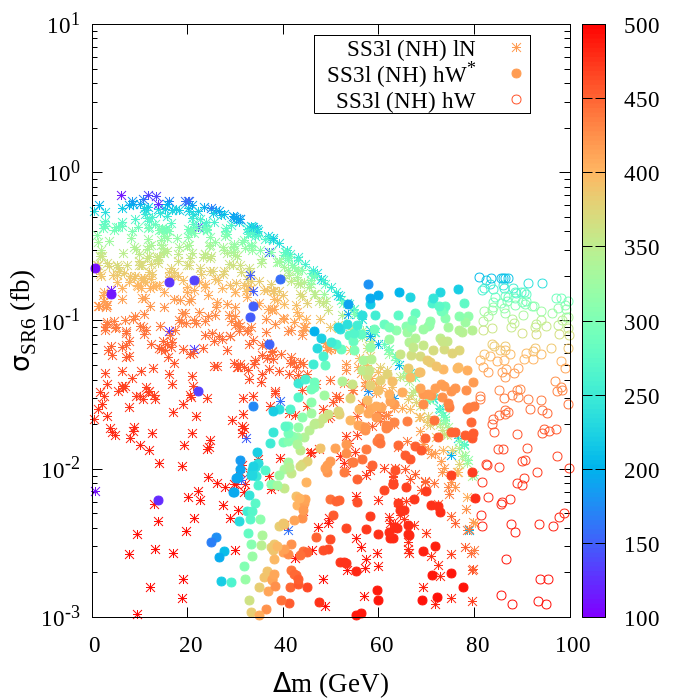} 
\includegraphics[height=5.4cm,width=5.4cm]{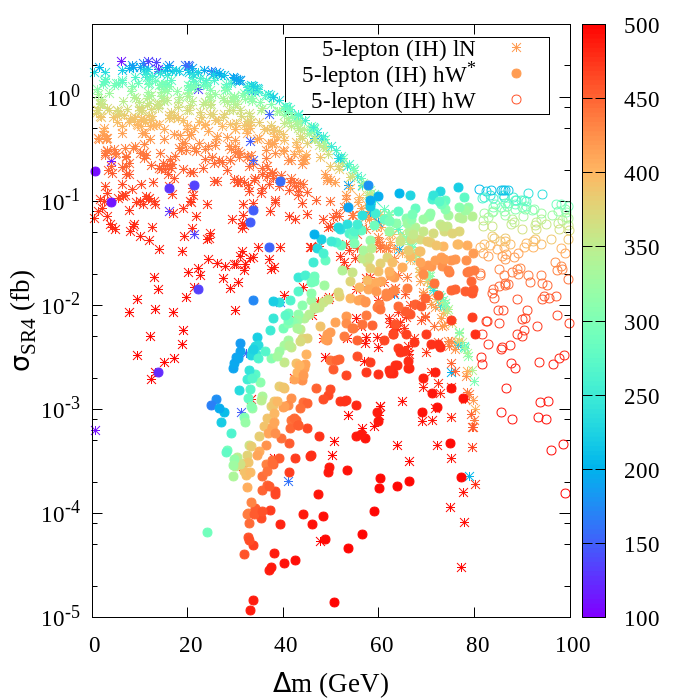} 
\includegraphics[height=5.4cm,width=5.4cm]{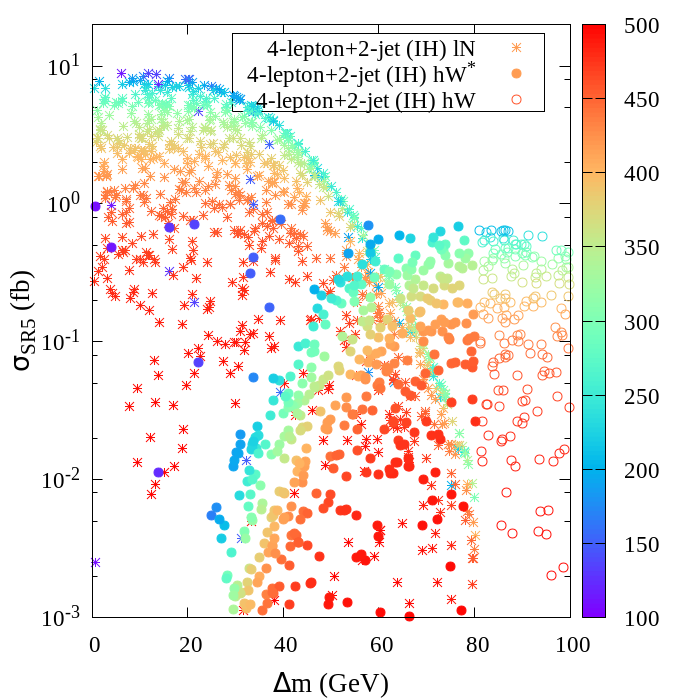} 
\includegraphics[height=5.4cm,width=5.4cm]{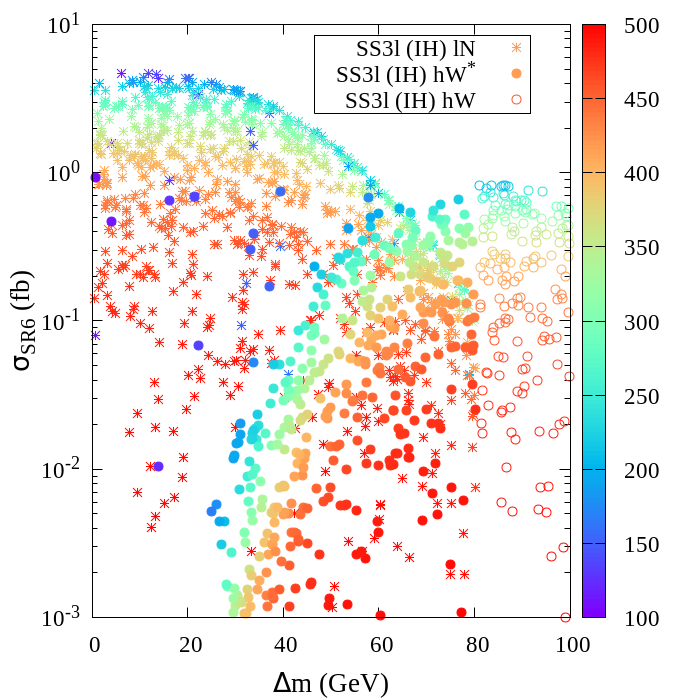} 
\caption{Variation of cross-sections corresponding to the signal regions {\bf SR4}, {\bf SR5} and {\bf SR6} as a function of $\Delta m$ at an $e^+e^-$ collider 
with 1 TeV center-of-mass energy. The two rows represent scenarios with normal and inverted hierarchy of the light neutrino masses respectively.
The color coding represents variation of $\mch$. For all points, heavy neutrino masses are kept at 100 GeV.}
\label{fig:SR_var_ee}
\end{center}
\end{figure}
The corresponding signal rates are showcased as a function of $\Delta m$ in Fig.~\ref{fig:SR_var_ee}. 
Trends of the distributions are similar to what we obtained for the LHC case. However, the difference in the production cross-section 
is manifested by the signal cross-sections indicating a larger event rate at the LHC for similar final states at low $\mch$ 
region. The rapid fall in production cross-section with increasing $\mch$ at the LHC makes it less relevant for heavier charged Higgs 
masses. An $e^+e^-$ collider can be more effective provided the center-of-mass energy is large enough for the production. Here, the 
signal rates drop alarmingly close to $\mch\sim 500~{\rm GeV}$ due to the choice of center-of-mass energy as 1 TeV.   
\subsection{Analysis}
\label{sec:coll_sim}
In order to carry out the simulation, events were generated at the parton level 
using \texttt{MadGraph5} \cite{Alwall:2011uj,Alwall:2014hca} with \texttt{nn23lo1} parton distribution function \cite{Ball:2012cx,Ball:2014uwa} 
and the default dynamic factorisation and 
renormalisation scales \cite{madgraph_scale}. We have used \texttt{PYTHIA} \cite{Sjostrand:2006za} for the subsequent decay of the particles, showering and hadronization. 
After that the events are passed through \texttt{Delphes} \cite{deFavereau:2013fsa,Selvaggi:2014mya,Mertens:2015kba} for detector simulation. 
Jets have been reconstructed using 
{\it anti-kT} algorithm via \texttt{FastJet} \cite{Cacciari:2011ma,Cacciari:2008gp}. The $b$-jet and $\tau$-jet tagging efficiencies as well as 
the mistagging efficiencies of the light jets as $b$ or $\tau$-jet have been incorporated according to the latest ATLAS studies in this regard \cite{ATLAS:2015-022}.   
\subsubsection*{Primary selection criteria}
\label{sec:selcn} 
We have applied the following cuts ({\bf C0}) on the jets, leptons and photons in order to identify them as final state particles: 
\begin{itemize}
\item All the charged leptons are selected with a transverse momentum threshold $p_T^{\ell} > 10$ GeV and in the pseudo-rapidity 
window $|\eta|^{\ell} < 2.5$. 
\item All the jets including $b$-jets and $\tau$-jets must have $p_T^{j} > 20$ GeV and $|\eta|^{j} < 2.5$.
\item We demand $\Delta R_{ij} > 0.4$ between all possible pairs of the final state particles to make sure they are well 
separated.
\end{itemize}
As discussed in Section~\ref{sec:cons_bp}, the choice of neutrino mass hierarchy affects the branching ratios 
of the neutrinophilic Higgs as well as the heavy neutrinos in certain flavor specific decay modes. 
Thus,
the hierarchical effect is reflected 
by the abundance of certain flavor of leptons in the signal events. As we have seen, 
one would expect less abundance of electrons in the final states for normal hierarchy 
scenario compared to that for inverted hierarchy. 
\begin{figure}[h!]
\begin{center}
\includegraphics[scale=0.45]{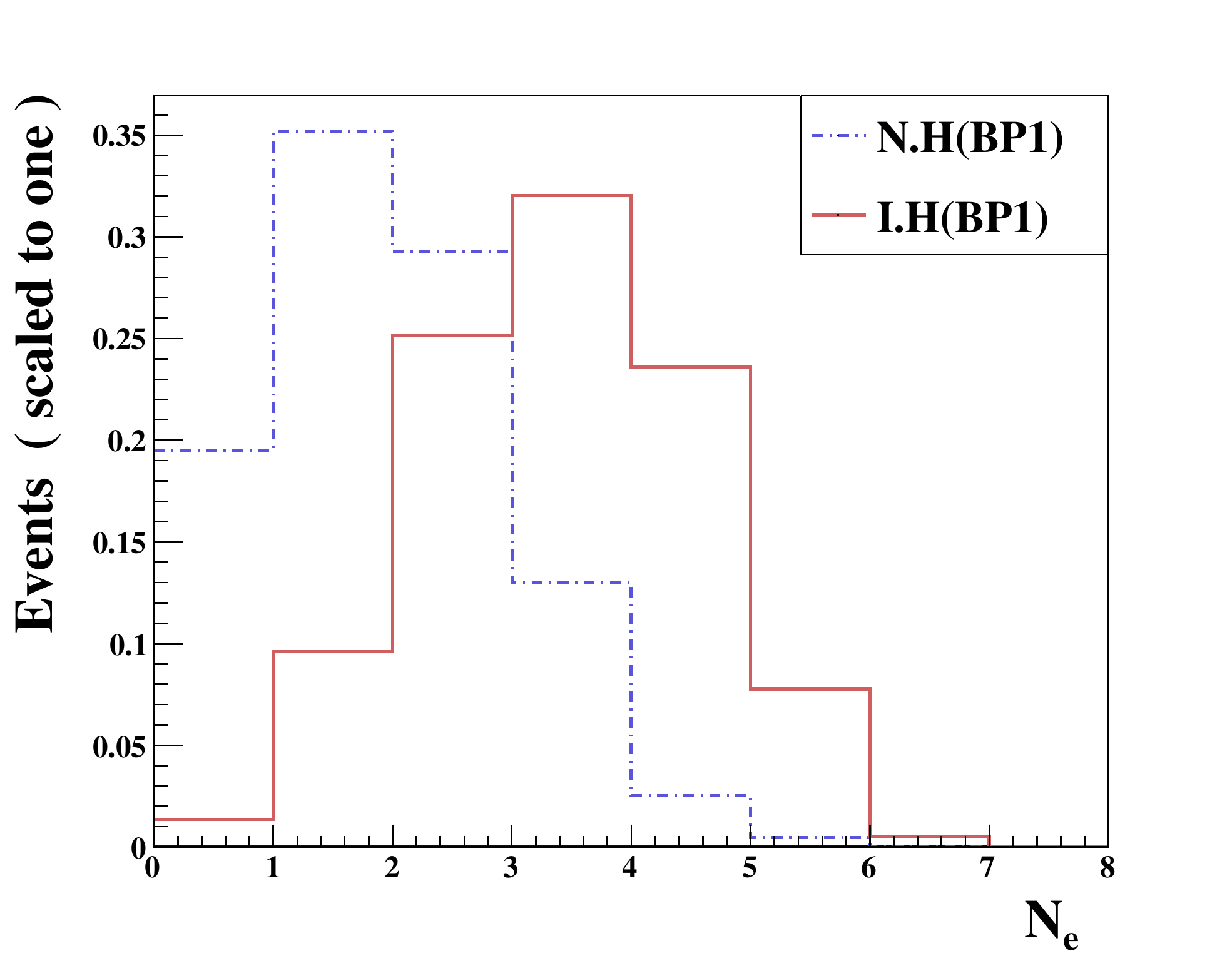} 
\caption{Electron multiplicity distribution for BP1 in normal and inverted hierarchy scenarios indicated by blue 
and red lines respectively. The distributions correspond to the choice of final states as per {\bf SR2}.}
\label{fig:bp1_nelec}
\end{center}
\end{figure}
This feature is evident in Fig.~\ref{fig:bp1_nelec} which shows the electron multiplicity in the final state with 
at least four leptons for {\bf BP1} in normal as well as inverted hierarchy scenarios. Such lepton multiplicity distributions 
can thus provide indirect probe of existing neutrino mass hierarchy.     
\subsection{Results@LHC13}
\label{sec:lhc13}
In the context of LHC, we have studied the final states corresponding to {\bf SR1}, {\bf SR2} and {\bf SR3} as defined in 
Section~\ref{sec:id_sr}. Although the choice of our signal regions ensure small or no SM background, we have checked the relevant 
production channels, $t\bar t$, $t\bar tZ(\gamma^*)$, $t\bar tW$, $WWZ$, $WZZ$, $ZZZ$ and
$ZZ$ + jets nevertheless in this regard.
In Table~\ref{tab:bp_sig_lhc} we show the expected number of different signal  
events at 13 TeV run of the LHC with an integrated luminosity ($\mathcal{L}$) of 1000 $\ifb$ after imposing a transverse missing 
energy cut, $\met > 50$ GeV and $b$-jet veto ({\bf C1}) \footnote{These cuts help reduce some of the surviving SM background contributions.}, 
in addition to the primary selection criteria, {\bf C0}. The choice of our signal regions combined with the cuts 
{\bf C1} render the SM backgrounds to negligible event numbers. We have 
observed that our {\bf SR1} is nearly backgroundless whereas {\bf SR2} and 
{\bf SR3} are left with 2 and 1 SM-background events, respectively, at 
1000 $\ifb$ integrated luminosity. As for the obtained signal event numbers, one can easily get an estimate of the expected 
rate from Fig.~\ref{fig:SR_var_lhc} for the different final states. However, note that in these figures the heavy neutrino mass is 
kept fixed at 100 GeV and if this mass is changed, so does the heavy neutrino branching ratios and hence the signal cross-sections. 
However, the cross-sections shown in these figures are good enough for order of magnitude estimation for a given $\mch$. 
\begin{table}[h!]
\begin{center}
\begin{tabular}{||c||c||c||c|c|c||} 
\hline\hline 
\multicolumn{1}{||c||}{Benchmark } &
\multicolumn{1}{|c||}{Production } &
\multicolumn{1}{|c||}{Neutrino } &
\multicolumn{3}{|c||}{Number of Events } \\
\multicolumn{1}{||c||}{Points} &
\multicolumn{1}{|c||}{cross-section (fb) } &
\multicolumn{1}{|c||}{ hierarchy} &
\multicolumn{3}{|c||}{ ($\mathcal{L}=$ 1000 $\ifb$)} \\
\cline{4-6}
  & ($\sqrt{s}=13$ TeV) & & \parbox[c]{1.5cm}{\bf{\hspace{0.1cm}SR1}}  & \parbox[c]{1.5cm}{\bf{\hspace{0.1cm}SR2}} & \parbox[c]{1.5cm}{\bf{\hspace{0.1cm}SR3}} \\ 
\hline\hline 
\multicolumn{1}{||c||}{{\bf BP1}} &
\multicolumn{1}{|c||}{60.71} &
\multicolumn{1}{|c||}{ Normal} &
\multicolumn{1}{|c|}{8} &
\multicolumn{1}{|c|}{130} &
\multicolumn{1}{|c||}{247}\\
\cline{3-6}
\multicolumn{1}{||c||}{} &
\multicolumn{1}{|c||}{ } &
\multicolumn{1}{|c||}{ Inverted} &
\multicolumn{1}{|c|}{25} &
\multicolumn{1}{|c|}{343} &
\multicolumn{1}{|c||}{397} \\
\hline
\multicolumn{1}{||c||}{{\bf BP2}} &
\multicolumn{1}{|c||}{22.13} &
\multicolumn{1}{|c||}{ Normal} &
\multicolumn{1}{|c|}{-} &
\multicolumn{1}{|c|}{13} &
\multicolumn{1}{|c||}{42} \\
\cline{3-6}
\multicolumn{1}{||c||}{} &
\multicolumn{1}{|c||}{ } &
\multicolumn{1}{|c||}{ Inverted} &
\multicolumn{1}{|c|}{1} &
\multicolumn{1}{|c|}{24} &
\multicolumn{1}{|c||}{55} \\
\hline
\multicolumn{1}{||c||}{{\bf BP3}} &
\multicolumn{1}{|c||}{6.72} &
\multicolumn{1}{|c||}{ Normal} &
\multicolumn{1}{|c|}{3} &
\multicolumn{1}{|c|}{40} &
\multicolumn{1}{|c||}{67} \\
\cline{3-6}
\multicolumn{1}{||c||}{} &
\multicolumn{1}{|c||}{ } &
\multicolumn{1}{|c||}{ Inverted} &
\multicolumn{1}{|c|}{8} &
\multicolumn{1}{|c|}{86} &
\multicolumn{1}{|c||}{101} \\
\hline
\multicolumn{1}{||c||}{{\bf BP4}} &
\multicolumn{1}{|c||}{27.34} &
\multicolumn{1}{|c||}{ Normal} &
\multicolumn{1}{|c|}{1} &
\multicolumn{1}{|c|}{26} &
\multicolumn{1}{|c||}{71} \\
\cline{3-6}
\multicolumn{1}{||c||}{} &
\multicolumn{1}{|c||}{ } &
\multicolumn{1}{|c||}{ Inverted} &
\multicolumn{1}{|c|}{3} &
\multicolumn{1}{|c|}{60} &
\multicolumn{1}{|c||}{112} \\
\hline\hline
\end{tabular}
\caption{Charged Higgs pair production cross-sections and number of events corresponding to the three different signal regions 
at 1000 $\ifb$ luminosity at 13 TeV LHC for our chosen benchmark points.}
\label{tab:bp_sig_lhc}
\end{center}
\end{table} 

As expected {\bf SR1} has the smallest event rate owing to its large lepton multiplicity, but with negligible SM background. 
Thus it can be a very clean signal but only if the charged 
Higgs mass is on the lighter side, as in {\bf BP1}, and at least one of the heavy neutrinos is lighter than the charged Higgs.  
The situation, however, worsens considerably with increasing charged Higgs mass, as indicated by {\bf BP3}. 
{\bf SR2} has a much better event rate and can probe {\bf BP1} and {\bf BP3} at much lower luminosity 
than {\bf SR1}. As the numbers in Table~\ref{tab:bp_sig_lhc} indicate, the inverted hierarchy scenario for {\bf BP1} 
can be probed with a 3$\sigma$ statistical significance at an integrated luminosity of $\sim$ 30 $\ifb$ in both {\bf SR2} and {\bf SR3}, i.e, 
if these signal regions are studied, $\mch$ in this mass range can be probed and possibly be excluded with the LHC data already accumulated. 
For the corresponding normal hierarchy 
case however, for the same benchmark point, one needs ${\mathcal L}\sim 50~\ifb$ for similar discovery significance in {\bf SR3}. 
{\bf BP3} requires an integrated luminosity of $\sim 100~{\ifb}$ (for inverted hierarchy) or more.
For the benchmark points like {\bf BP4} and {\bf BP2}, the decay $\ch\to\ell\hn$ is either suppressed or absent altogether. 
Thus for such points {\bf SR1} ceases to be a viable signal region while {\bf SR2} is relevant only at large luminosities.  
In this case {\bf SR3} turns out to be the most viable signal region. In this signal region, to achieve 3$\sigma$ statistical 
significance in the inverted hierarchy case of {\bf BP4} and {\bf BP2} one requires ${\mathcal L}\sim 100~\ifb$ and $\sim 200~\ifb$ 
respectively. For all the benchmark points, the choice of 
light neutrino mass hierarchy is clearly manifested through the different signal event rates. Evidently, with multi-leptonic
final states, an inverted hierarchy scenario is more likely to be probed at lower luminosities at the LHC.
\subsection{Results@$e^+e^-$ collider}
\label{sec:epem1T}
In Table~\ref{tab:bp_sig_ee} below we have presented the expected number of different signal  
events at 1 TeV run of an $e^+e^-$ collider with an integrated luminosity of 100 $\ifb$ after 
imposing a missing energy cut, $\me > 50~{\rm GeV}$ GeV and b-jet veto ({\bf D1}), 
in addition to the primary selection criteria, {\bf C0}.  
The event rates are quite good and devoid of any direct SM background, which makes it an ideal platform 
to look for a neutrinophilic charged Higgs. 
Although the number of events shown in Table~\ref{tab:bp_sig_ee} correspond to $\mathcal{L}=$ 100 $\ifb$, 
the inverted hierarchy scenarios in {\bf BP1} and {\bf BP3} can be probed with a statistical significance 
of 3$\sigma$ at a much lower luminosity ($\sim$ 10 $\ifb$). Note the improved event rates in signal regions 
{\bf SR5} and {\bf SR6} despite of the smaller $\ch$ pair production cross-section in {\bf BP3} over those of {\bf BP1}. 
This is a consequence of increased hadronic 
branching ratio of $N$ and improved cut efficiency due to the larger mass gap between $\ch$ and $N$ in {\bf BP3}. Even {\bf BP2} which can be probed at the LHC 
only at very high luminosity can be probed here at around $\mathcal{L}=50~\ifb$ with similar statistical significance 
via {\bf SR5} which turns out to be the 
most favored signal in general for all the benchmark points. The overall signal rate is relatively weaker 
in {\bf SR6} due to better lepton tagging efficiency at a lepton collider, which results in smaller number of
events with exactly three same-sign leptons as demanded.      
\begin{table}[h!]
\begin{center}
\begin{tabular}{||c||c||c||c|c|c||}
\hline\hline
\multicolumn{1}{||c||}{Benchmark } &
\multicolumn{1}{|c||}{Production } &
\multicolumn{1}{|c||}{Neutrino } &
\multicolumn{3}{|c||}{Number of Events } \\
\multicolumn{1}{||c||}{Points} &
\multicolumn{1}{|c||}{cross-section (fb) } &
\multicolumn{1}{|c||}{ hierarchy} &
\multicolumn{3}{|c||}{ ($\mathcal{L}=$ 100 $\ifb$)} \\
\cline{4-6}
  & ($\sqrt{s}=1$ TeV) & & \parbox[c]{1.5cm}{\bf{\hspace{0.1cm}SR4}}  & \parbox[c]{1.5cm}{\bf{\hspace{0.1cm}SR5}} & \parbox[c]{1.5cm}{\bf{\hspace{0.1cm}SR6}} \\
\hline\hline
\multicolumn{1}{||c||}{{\bf BP1}} &
\multicolumn{1}{|c||}{22.83} &
\multicolumn{1}{|c||}{ Normal} &
\multicolumn{1}{|c|}{36} &
\multicolumn{1}{|c|}{47} &
\multicolumn{1}{|c||}{9} \\
\cline{3-6}
\multicolumn{1}{||c||}{} &
\multicolumn{1}{|c||}{ } &
\multicolumn{1}{|c||}{ Inverted} &
\multicolumn{1}{|c|}{77} &
\multicolumn{1}{|c|}{90} &
\multicolumn{1}{|c||}{9} \\
\hline
\multicolumn{1}{||c||}{{\bf BP2}} &
\multicolumn{1}{|c||}{16.91} &
\multicolumn{1}{|c||}{ Normal} &
\multicolumn{1}{|c|}{5} &
\multicolumn{1}{|c|}{16} &
\multicolumn{1}{|c||}{6} \\
\cline{3-6}
\multicolumn{1}{||c||}{} &
\multicolumn{1}{|c||}{ } &
\multicolumn{1}{|c||}{ Inverted} &
\multicolumn{1}{|c|}{6} &
\multicolumn{1}{|c|}{23} &
\multicolumn{1}{|c||}{8} \\
\hline
\multicolumn{1}{||c||}{{\bf BP3}} &
\multicolumn{1}{|c||}{12.48} &
\multicolumn{1}{|c||}{ Normal} &
\multicolumn{1}{|c|}{30} &
\multicolumn{1}{|c|}{75} &
\multicolumn{1}{|c||}{16} \\
\cline{3-6}
\multicolumn{1}{||c||}{} &
\multicolumn{1}{|c||}{ } &
\multicolumn{1}{|c||}{ Inverted} &
\multicolumn{1}{|c|}{63} &
\multicolumn{1}{|c|}{122} &
\multicolumn{1}{|c||}{17} \\
\hline
\multicolumn{1}{||c||}{{\bf BP4}} &
\multicolumn{1}{|c||}{18.44} &
\multicolumn{1}{|c||}{ Normal} &
\multicolumn{1}{|c|}{8} &
\multicolumn{1}{|c|}{22} &
\multicolumn{1}{|c||}{8} \\
\cline{3-6}
\multicolumn{1}{||c||}{} &
\multicolumn{1}{|c||}{ } &
\multicolumn{1}{|c||}{ Inverted} &
\multicolumn{1}{|c|}{16} &
\multicolumn{1}{|c|}{39} &
\multicolumn{1}{|c||}{12} \\
\hline\hline
\end{tabular}
\caption{Charged Higgs pair production cross-sections and number of events corresponding to the three different signal regions
at 100 $\ifb$ luminosity at 1 TeV $e^+e^-$ collider for our chosen benchmark points.}
\label{tab:bp_sig_ee}
\end{center}
\end{table}
\section{Summary and Conclusions}
\label{sec:concl}
We have considered a simple extension of the SM with one additional scalar doublet and three generations of 
singlet right-handed Majorana neutrinos, where the additional Higgs states interact with the SM sector only via the 
right-handed neutrinos. The model, known as the neutrinophilic Higgs doublet model, is a well-motivated framework from 
the viewpoint of neutrino mass generation. The light neutrinos gain tiny non-zero masses via Type-I seesaw mechanism 
when the neutrinophilic Higgs obtains a VEV to break the ${\mathcal Z}_2$ symmetry. We have discussed in brief why 
the spontaneous breaking of 
the ${\mathcal Z}_2$ symmetry is disfavored, if one imposes the constraints derived from the CMB temperature anisotropies 
induced by domain walls as well as LFV decay branching ratios. We have, therefore, considered a scenario where the 
${\mathcal Z}_2$ parity is broken explicitly and thus is devoid of the domain wall problem. In such a scenario, the charged 
Higgs can have interesting collider phenomenology, explored in this work. Depending on the different 
decay modes of the neutrinophilic charged Higgs, we have identified some particularly clean signal regions likely 
to provide a hint of $\nu$HDM scenarios at the collider experiments. 

We have also highlighted the interesting roleplay of the light neutrino 
mass hierarchy. Whether the neutrinos follow normal or inverted hierarchy, is likely to be manifested via multiplicity of 
different flavored leptons in the final state. Thus such a finding at the collider experiments can complement 
the neutrino oscillation experiments which are yet to ascertain the correct mass hierarchy of the three light neutrinos.

The fact that the charged Higgs pair production cross-section falls quite rapidly at the LHC with increasing mass, led us to
perform a comparative study between the LHC and a future $e^+e^-$ machine in order to probe such scenarios. We observed that 
although LHC is quite efficient to probe light charged Higgs masses, an $e^+e^-$ collider will be able to probe 
a much larger parameter space with heavier states.
\section{Acknowledgements}
KH and SM acknowledge the H2020-MSCA-RICE-2014 grant no. 645722 (NonMinimalHiggs).
TK expresses his gratitude to Magnus Ehrnrooth foundation for financial support. 
The work of SKR was partially supported by funding available from the Department of Atomic 
Energy, Government of India, for the Regional Centre for Accelerator-based Particle Physics
(RECAPP), Harish-Chandra Research Institute. 

\section{Appendix}
\subsection{Relevant interaction vertices}
 \begin{align*}
& \ell^{\mp i}\hn^jH_{\nu}^{\pm k} && i\sum_{a=1}^{3}\sum_{b=1}^3 y_{\nu}^{ab}U_L^{ia}U_N^{j3+b} & \\
& \hh^iH_{\nu}^{\pm j} W^{\mp} && \frac{i}{2}gZ^{i2}(p^{\hh^i} - p^{H_{\nu}^{\pm j}})_{\mu} & \\
& \nu^i\hn^j\hh^k & - & i\sum_{a=1}^{3}\sum_{b=1}^{3}U_N^{ia}U_N^{j3+b}y_{\nu}^{ab}Z^{k2} &
\end{align*}
where $U_L^{ij}$, $U_N^{ij}$ and $Z^{ij}$ are the charged lepton, neutrino and CP-even neutral Higgs
mixing matrices, the bases of the mass matrices being $\{e, \mu, \tau\}$, $\{\nu_e, \nu_{\mu}, \nu_{\tau}, N^c_e, N^c_{\mu}, N^c_{\tau} \}$
and $\{\phi, \phi_{\nu} \}$ respectively. Note that, $U_L^{ij}$ is a diagonal matrix.
\subsection{Lepton Flavor Violating Branching Ratios}
\begin{table}[ht!]
\begin{center}
\begin{tabular}{||c||c|c||c|c||c|c||c|c||}
\hline
\multicolumn{1}{||c||}{LFV} &
\multicolumn{2}{|c||}{\bf BP1} &
\multicolumn{2}{|c||}{\bf BP2} &
\multicolumn{2}{|c||}{\bf BP3} &
\multicolumn{2}{|c||}{\bf BP4} \\
\cline{2-9}
Process & Normal & Inverted & Normal & Inverted & Normal & Inverted & Normal & Inverted \\
\hline\hline
BR($\mu\to e\gamma$)($\times 10^{15}$) & 2.97  & 0.79  & 0.56  & 0.21 & 0.81 & 0.31 & 0.99 & 0.38 \\
BR($\tau\to e\gamma$)($\times 10^{16}$) & 0.38  & 2.05  & 0.10 & 0.56 & 0.15 & 0.80 & 0.18 & 0.98 \\
BR($\tau\to\mu\gamma$)($\times 10^{15}$) & 2.90  & 3.87  & 0.79 & 1.05 & 1.14 & 1.52 & 1.39 & 1.86 \\
BR($\mu\to e e e$)($\times 10^{17}$) & 1.72  & 0.65  & 0.46 & 0.18 & 0.68 & 0.26 & 0.82 & 0.31 \\
BR($\tau\to e e e$)($\times 10^{18}$) & 0.51  & 2.73  & 0.14 & 0.73 & 0.20 & 1.07 & 0.24 & 1.30 \\
BR($\tau\to\mu\mu\mu$)($\times 10^{17}$) & 1.12  & 1.50  & 0.30 & 0.39 & 0.45 & 0.60 & 0.53 & 0.71 \\
\hline\hline
\end{tabular}
\caption{Lepton flavor violating branching ratios obtained for our chosen benchmark points.}
\label{tab:lfv_br}
\end{center}
\end{table}
In Table~\ref{tab:lfv_br} we have shown the obtained branching ratios for various lepton flavor violating processes
corresponding to the four benchmark points. BR($\mu\to e\gamma$) is projected to be probed experimentally up to
$6.0\times 10^{-14}$ in near future \cite{Baldini:2013ke}.
As indicated by the numbers, the obtained branching ratios for this process are at least one order of magnitude smaller for our benchmark points.
The rest of these obtained LFV branching ratios are several orders of magnitude below the present experimental sensitivity in the respective channels.
\newpage
\bibliography{nuphil_hig}{}
\end{document}